\documentclass[mreferee]{gji}
\usepackage{timet}
\usepackage[dvips]{graphicx}
\usepackage{amsmath}
\usepackage{color}
\usepackage{url}
\makeatletter
\let\bibhang\@undefined
\makeatother
\usepackage[authoryear]{natbib}
\usepackage{algorithm}
\usepackage{algorithmic}
\usepackage{bm}
\usepackage{txfonts}
\usepackage{cases}
\usepackage{multirow}
\usepackage{multicol}
\usepackage{lineno}
\usepackage{amssymb}

\newcommand{\argmin}{\mathop{\rm arg~min}\limits}

\title[Bayesian 3D travel-time tomography with PINNs]{Bayesian three-dimensional seismic travel-time tomography for active- and passive-source seismic data using physics-informed neural networks}

\author[R.~Agata et al.]
  {Ryoichiro Agata$^1$, Kazuya Shiraishi$^1$, Gou Fujie$^1$ and Dan Bassett$^2$\\
  $^1$ Japan Agency for Marine-Earth Science and Technology\\
  $^2$ Earth Science New Zealand}

\date{Accepted xxx. Received xxx; in original form xxx}
\pagerange{\pageref{firstpage}--\pageref{lastpage}}
\volume{000}
\pubyear{2026}

\begin{document}
\label{firstpage}
\maketitle

\begin{summary}
Accurate three-dimensional (3D) seismic velocity modeling through seismic travel-time tomography using both active- and passive-source data provides critical underpinning models for seismicity monitoring and hazard assessment. Because travel-time tomography is an inherently ill-posed inverse problem, uncertainty quantification (UQ) of the estimated models using Bayesian methods is also important for reliable downstream interpretations and analyses. However, Bayesian inference for 3D tomography based on conventional grid-based representations faces the ``curse of dimensionality'' and severe computational bottlenecks. 
Consequently, rigorous Bayesian UQ for margin-wide 3D travel-time tomography has remained largely unexplored, despite its importance for reliable interpretation of large-scale velocity models.
In this study, we propose a meshless 3D Bayesian travel-time tomography method that combines physics-informed neural networks (PINNs) with a neural representation of the velocity structure, enabling tractable and data-efficient Bayesian inference through function-space Particle-based Variational Inference (fParVI). 
To efficiently integrate passive-source data into the Bayesian estimation of the velocity structure, we introduce an analytical marginalization strategy that treats uncertain parameters related to the origin time and location as nuisance parameters, with passive-source relocation carried out in post-processing. 
We validated the capability of our approach for 3D problems through synthetic experiments. These experiments demonstrate its robustness against biased prior source locations compared to standard methods that suffer from overfitting. Furthermore, we applied the method to a real-world dataset off the Kii Peninsula, Nankai Trough, using over 380,000 travel-time picks from marine active-source surveys and natural earthquakes. Our probabilistic 3D ensemble successfully resolves key geological features, such as the Kumano Pluton, and provides data-consistent uncertainty maps that offer new quantitative insights into the reliability of previously debated deep low-velocity zones. 
The posterior mean hypocenters shifted mainly in the vertical direction by 10-15 km, consistent with the deterministic relocation results of a previous study.
Finally, the continuous neural representation drastically reduces storage requirements for the entire ensemble velocity model, highlighting the scalability and data efficiency of the proposed framework for Bayesian seismic travel-time tomography.
\end{summary}


\section{Introduction}

Understanding the three-dimensional (3D) seismic velocity and elastic structure of subduction zones is critical for accurate seismicity monitoring and hazard assessment.  
Seismic tomography remains a primary tool for retrieving such subsurface information.  
In subduction zones, passive source tomography utilizing numerous local earthquakes is widely applicable.  
Furthermore, in regions where potentially damaging earthquakes can originate, active source seismic surveys are often conducted to probe subsurface structures.  
These two data types are complementary: local passive-source data generally resolve deeper structures, whereas active-source data provide high-resolution images of shallower depths.  
Integrating both datasets is, therefore, effective for obtaining a comprehensive 3D P-wave velocity model from the shallow crust down to greater depths.  
Consequently, beyond individual analyses, recent efforts have focused on combining these datasets to construct unified 3D velocity structure models \cite[]{Arnulf2022NatGeo,Bassett2022JGR-SE,Yamamoto2022PEPS,Bassett2025JGR-SE}.

Velocity structure estimation is inherently an ill-posed inverse problem due to noise and the sparsity of observation data.  
To ensure the reliability for downstream interpretations, it is crucial to also be able to quantify the uncertainty of estimated velocity models.  
Bayesian inference has played a central role in uncertainty quantification (UQ) in geophysical studies \cite[]{Mosegaard1995JGR,Tarantola2005Textbook}.  
For travel-time tomographic problems, which are highly nonlinear, ensemble-based Bayesian methods have been proposed to represent uncertainty through sampling methods \cite[e.g.,][]{Bodin2009,Piana2015} or particle-based variational inference methods \cite[e.g.,][]{Zhang2020Seismic,Yang2025GJI}.  
However, when targeting broad 3D velocity structures, the parameter space becomes high-dimensional, leading to the ``curse of dimensionality'', and Bayesian inference becomes practically intractable.  
Handling an ensemble of massive volumes of data is also challenging.  
An emerging approach is to represent the 3D velocity field using a neural network, which can be viewed as a scientific application of (implicit) neural representations \cite[]{Sitzmann2020NeurIPS}, popularized in computer vision and graphics as coordinate-based neural field representations.  
Neural representations of subsurface velocity structures have been gaining attention in full waveform inversion, primarily in the context of reducing the dependence of the results on the initial model \cite[]{Sun2023JGR-SE} and also for Bayesian inference \cite[]{Zhang2023IEEE-TGRS}.  
Neural representations provide a continuous, mesh-free parameterization whose degrees of freedom are decoupled from grid resolution.  
This mitigates the exponential growth in parameters and storage/compute costs that arises when refining a grid, and can therefore ease some practical aspects of the curse of dimensionality in grid-based representations.  
In this context of mesh-free parameterization, physics-informed neural networks (PINN) \cite[]{Raissi2019}, which approximate the solution of partial differential equations (PDEs) using NNs, have gained attention as mesh-free PDE solvers.  
A fully mesh-free travel-time tomography method combining PINNs with an NN-based velocity representation has been proposed \cite[]{Waheed2021PINNeik,Chen2022JGR}.  
That approach has recently been extended to Bayesian inference \cite[]{Agata2023} and applied to 2D $P$-wave velocity tomography along a seismic survey line off the Kii Peninsula in Southwest Japan \cite[]{Agata2025SR}.  
However, Bayesian tomography based on neural representations and PINNs has so far been limited to 2D profiles beneath seismic survey lines using active sources, and its applicability to 3D model estimation remains unexplored.

When integrating local passive earthquakes into such a 3D Bayesian tomography framework, a practical complication arises from the uncertainty in passive-source parameters, namely, location and origin time.  
Unlike active sources, whose locations and origin times are well constrained, the hypocenters and origin times of local passive sources derived from routine catalogs are uncertain and velocity model dependent.  
Using these parameters directly in tomography can introduce artifacts into the velocity structure.  
In standard deterministic tomography, this issue is typically addressed by simultaneously inverting for source locations and velocity structure, often using differential travel times \cite[e.g.,][]{Zhang2003}.  
However, treating local passive sources, which can number in the tens of thousands, as unknown parameters alongside the velocity structure in a Bayesian framework results in a high-dimensional multi-scale problem that is computationally prohibitive.  
Although some studies have achieved this by using relatively low-dimensional parameterizations for the velocity structure \cite[]{Piana2015,Yang2025GJI}, such examples are limited.  
A practical approach to address the uncertainty of local passive-source parameters is to treat them as nuisance parameters and account for their effects through posterior marginalization, making the Bayesian estimation focus solely on the velocity structure.  
This strategy has been employed in subsurface studies with several different algorithms \cite[]{Duputel2014,Gesret2015,Agata2021} (note that, conversely, those studies treated velocity as a nuisance parameter for source estimation) and fundamentally relies on analytical marginalization based on Gaussian assumptions and linearization, as described in classical textbooks \cite[e.g.,][]{Tarantola2005Textbook}.  
However, passive-source parameters are not truly ``nuisance'' parameters; rather, they are often also parameters of interest.  
\cite{Agata2021} showed that the posterior distribution of parameters eliminated by marginalization can be obtained through a post-processing step using the samples of the main estimated parameters.  
Such a two-step approach is expected to make the Bayesian estimation of the velocity structure tractable in the first step while relocating the passive sources in the second step.

The objective of this study is to perform PINN-based 3D travel-time tomography using both active and local passive sources.  
Our primary contribution is to extend the PINN-based Bayesian tomography method described in \cite{Agata2023} to 3D model estimation and to evaluate its applicability in realistic settings.  
To the best of our knowledge, this is one of the first attempts to realize Bayesian UQ for 3D travel-time tomography at the scale of a subduction-zone segment.
In addition, we propose a marginalization strategy that enables tractable Bayesian estimation of the velocity structure, while also enabling the relocation of passive sources as a post-processing step.  
We demonstrate the proposed approach through numerical experiments and apply it to actual data from active surveys and natural earthquakes in the Nankai Trough region, Southwest Japan.


\section{Methods}
\label{sct:methods}
\subsection{PINN-based Bayesian seismic tomography for well known source locations}
\label{sct:fParVI}

We first briefly describe the PINN-based Bayesian seismic tomography method proposed by \cite{Agata2023}, assuming the use of active-source data with well-known source parameters.  
We introduce NNs to represent the velocity structure and the travel-time function, which are defined as
\begin{eqnarray}
v(\mathbf{x}) &\simeq& f_{v}(\mathbf{x},\boldsymbol{\theta}_{v})\nonumber\\
T(\mathbf{x},\mathbf{x}_{s}) &\simeq& f_{T}(\mathbf{x},\mathbf{x}_{s},\boldsymbol{\theta}_{T})\nonumber,
\end{eqnarray}
where $f_{v}$ and $f_{T}$ are the NN-based functions, and $\boldsymbol{\theta}_{v}$ and $\boldsymbol{\theta}_{T}$ are the weight parameters of the NNs.  
For any updated velocity structure and $\boldsymbol{\theta}_{v}$, the travel-time NN is trained to satisfy the eikonal equation through the PINN framework \cite[]{Smith2021,Waheed2021PINNeik,Grubas2023}.  
Specifically, $\boldsymbol{\theta}_{T}$ is trained by minimizing a loss function composed of the residual of the eikonal equation 
\begin{equation}
|\nabla_{\mathbf{x}} f_T(\mathbf{x}, \mathbf{x}_s, \boldsymbol{\theta}_T)|^2 - 1/v^2(\mathbf{x}) = 0, \quad \forall \mathbf{x} \in \Omega,\label{eqn:eikonal}
\end{equation}
as
\begin{equation}
\boldsymbol{\theta}_{T} = \argmin_{\boldsymbol{\theta}_T} L(\boldsymbol{\theta}_T),\label{eqn:argmin_theta_T}
\end{equation}
where $L(\boldsymbol{\theta}_T)$ is the loss function defined as
\begin{equation}
L(\boldsymbol{\theta}_T) = \frac{1}{N_{\rm c}} \sum_{i=1}^{N_{\rm c}}\left(|\nabla_{\mathbf{x}} f_T(\mathbf{x}_{\rm c}^{(i)}, \mathbf{x}_{\rm s}^{(i)}, \boldsymbol{\theta}_T)|^2 - 1/f_v(\mathbf{x}_{\rm c}^{(i)}, \boldsymbol{\theta}_v)^2 \right)^2,\label{eqn:loss_eikonal}
\end{equation}
and $N_{\rm c}$ is the number of collocation points randomly sampled from the domain $\Omega$.  
$\mathbf{x}_{\rm c}$ and $\mathbf{x}_{\rm s}$ are the coordinates of the collocation and source points, respectively.  
The source condition $f_T(\mathbf{x}_s, \mathbf{x}_s, \boldsymbol{\theta}_T) = 0$ is imposed as a hard constraint through multiplicative factorization \cite[]{Waheed2021PINNeik}.  
After training, the PINN can predict travel time for any source and receiver locations in the domain $\Omega$.

We introduce an output vector ${\bf v}$ from the velocity NN at the query points $\mathbf{x}_{i}$,
\begin{eqnarray}
{\bf v} &=& \left[ f_v(\mathbf{x}_{i},\boldsymbol{\theta}_{v}) \right]_{i=1}^{N_v},
\end{eqnarray}
where $N_v$ is the number of query points.  
These query points can be randomly sampled from the domain $\Omega$.  
Similarly, the output vector for a set of observation points for the $i$-th source can be obtained as
\begin{eqnarray}
\mathbf{t}_{\mathrm{calc}}(\mathbf{x}_{\mathrm{s},i})= \left[ f_{T}({\bf x}_{i}^{j},\mathbf{x}_{\mathrm{s},i},\boldsymbol{\theta}_{T}) \right]_{j=1}^{N_{{\rm obs},i}},
\end{eqnarray}
where $N_{{\rm obs},i}$ is the number of observation points for the $i$-th source.

Based on Bayes' theorem, the posterior PDF of the velocity structure given the travel-time data between the sources and observation points is written as
\begin{eqnarray}
p({\bf v}|{\bf T}_{\rm obs}) &\propto& p({\bf T}_{\rm obs} | {\bf v}) p({\bf v}),\label{eqn:bayes}
\end{eqnarray}
where ${\bf v}$ is the velocity structure, ${\bf T}_{\rm obs}$ is the observed travel-time data, $p({\bf T}_{\rm obs} | {\bf v})$ is the likelihood function, and $p({\bf v})$ is the prior PDF of the velocity structure.  
Because ${\bf v}$ is produced as the output of a NN, $p({\bf v})$ should not be formulated as a finite-dimensional probability distribution over discretized grid-point parameters.  
Instead, it should be defined as a stochastic process over functions, such as a Gaussian process.  
The likelihood function includes the forward modeling of travel-time calculation for the given velocity structure.  
The data error is assumed to be independent for each source.  
Writing this explicitly, we have
\begin{eqnarray}
p({\bf T}_{\rm obs} | {\bf v}) = \prod_{i=1}^{N} p(\mathbf{t}_{\mathrm{obs},i} | \mathbf{t}_{\mathrm{calc}}(\mathbf{x}_{\mathrm{s},i})),
\end{eqnarray}
where $N$ is the number of sources.  
$\mathbf{t}_{\mathrm{obs},i}$ is the observed travel-time vector from the $i$-th source to the observation points.  
When $f_{T}$ is trained for ${\bf v}$ with the eikonal equation through the PINN framework, $\mathbf{t}_{\mathrm{calc}}(\mathbf{x}_{\mathrm{s},i})$ implicitly depends on ${\boldsymbol{\theta}_{v}}$ (see discussion later in this section).  
When the source parameters are well known, as in active-source tomography, the likelihood function is written as
\begin{eqnarray}
p(\mathbf{t}_{\mathrm{obs},i} | \mathbf{t}_{\mathrm{calc}}(\mathbf{x}_{\mathrm{s},i}))
= \frac{1}{\sqrt{(2\pi)^{n_i} \det(\mathbf{C}_{\mathrm{data}})}} \exp{\left(-\frac{1}{2} \Delta \mathbf{t}(\mathbf{x}_{\mathrm{s},i})^\top \mathbf{C}_{\mathrm{data}}^{-1}  \Delta \mathbf{t}(\mathbf{x}_{\mathrm{s},i})  \right)}, \label{eqn:likelihood}
\end{eqnarray} 
where $n_i$ is the number of observation points for the $i$-th source, $\mathbf{C}_{\mathrm{data}}$ is the covariance matrix of the data error, and $\Delta \mathbf{t}(\mathbf{x}_{\mathrm{s},i}) = \mathbf{t}_{\mathrm{obs},i} - \mathbf{t}_{\mathrm{calc}}(\mathbf{x}_{\mathrm{s},i})$.  
In the case of well-known source locations, we assume that the data error is independent for each source, i.e., $\mathbf{C}_{\mathrm{data}}$ is a diagonal matrix.  
Because probability densities, whose magnitudes vary significantly due to exponential functions, are not easy to handle numerically, log-posterior PDFs are often used instead in practice.  

When the gradient of the log-posterior PDF, $\nabla_{\bf v} \log p({\bf v}|{\bf T}_{\rm obs})$, is available, efficient sampling- or ensemble-based Bayesian inference methods, such as Hamiltonian Monte Carlo (HMC) \cite[]{Duane1987}, can be applied.  
We introduce one such approach, particle-based variational inference (ParVI) \cite[]{Liu2019ICML}, a class of Bayesian estimation methods that has been developed over the past decade and is known for its high efficiency and parallelism.  
ParVI iteratively transports a set of particles to approximate a target posterior distribution.  
The most widely used instance of ParVI is Stein variational gradient descent (SVGD) \cite[]{Liu2016}, which has also been gaining popularity in geophysical applications \cite[]{Zhang2020Seismic,Zhang2020Variational}.  
The iterative updating equation of SVGD applied to the velocity structure is given as follows:
\begin{eqnarray}
\boldsymbol{\phi}(\mathbf{v_i})=\frac{1}{n} \sum_{j=1}^{n} \lbrace \underset{\rm Driving\,force}{\underline{k(\mathbf{v}_{j}^{l},\mathbf{v_i}) \nabla_{\mathbf{v}_{j}^{l}}\log P(\mathbf{v}_{j}^{l}|{\bf d})}}+\underset{\rm Repulsive\,force}{\underline{\nabla_{\mathbf{v}_{j}^{l}} k(\mathbf{v}_{j}^{l},\mathbf{v_i})}} \rbrace,
\label{eqn:SVGD_vector}
\end{eqnarray}
where $k(\mathbf{x}, \cdot)$ represents a positive definite kernel, $i$ and $j$ are the indices of the particles, and $l$ is the iteration index.  
The ``driving force'' term is a smoothed gradient of the log-posterior density that moves the particles toward the high-density regions of the posterior PDF.  
The ``repulsive force'' term promotes diversity and prevents particles from concentrating around the mode of the target PDF.  
This combination of two forces results in an efficient nonparametric approximation of the posterior PDF with a finite number of particles.  
In the original SVGD, this update vector is applied to the set of parameters $\mathbf{v}_{i}$ as
\begin{eqnarray}    
\mathbf{v}_{i}^{l+1}=\mathbf{v}_{i}^{l}+\epsilon_{l} \boldsymbol{\phi}\left(\mathbf{v}_{i}^{l}\right),
\label{eqn:SVGD_update}
\end{eqnarray}
where $\epsilon_l$ is the step size at each iteration $l$.  
However, because the velocity structure in our study is represented by NNs, an update rule for the NN parameters $\boldsymbol{\theta}_{v}$ instead of one for the velocity structure ${\bf v}$ is required.  
\cite{Wang2019} proposed a variant of ParVI for NNs, called function-space ParVI (fParVI), which defines the ParVI update rule on the output function space of the NNs while applying the actual updates to the weight space of the NNs.  
This is simply achieved by using the Jacobian matrix of the NN output with respect to the weight parameters for parameter-space conversion.  
In the case of function-space SVGD (fSVGD), the parameters $\boldsymbol{\theta}_{v}$ of the $i$-th particle are updated as
\begin{eqnarray}
\boldsymbol{\theta}_{v,i}^{l+1}=\boldsymbol{\theta}_{v,i}^{l}+\epsilon_{l}
\frac{\partial {\bf v}_{i}}{\partial \boldsymbol{\theta}_{v,i}^{l}}^{\top}  \boldsymbol{\phi}\left(\mathbf{v}_{i}^{l+1}\right).
\label{eqn:SVGD_update_improved2}
\end{eqnarray}
Conducting Bayesian estimation involving NNs with ParVI in function space results in better estimation performance than the weight-space formulation because it forces ParVI to explore the PDF in function space, which is expected to be far less complex than the weight space \cite[]{Wang2019}.  
This approach is also capable of introducing a physically meaningful prior PDF for the velocity structure because Bayes' theorem is formulated in function space.  
Finally, using the optimized parameters, the target posterior PDF is approximated by a Dirac delta function as
\begin{eqnarray}
p(\mathbf{v} | {\bf T}_{\rm obs}) \approx \frac{1}{N} \sum_{i=1}^{N} \delta(\mathbf{v} - \mathbf{v}_{i}),
\end{eqnarray}
where $\mathbf{v}_i$ is the optimized velocity structure output from the velocity NN with $\boldsymbol{\theta}_{v,i}$.  
However, calculation of $\nabla_{\bf v} \log p({\bf v}|{\bf T}_{\rm obs})$ is not straightforward because the likelihood function, which fundamentally consists of the misfit between $\mathbf{t}_{\mathrm{calc}}$ and $\mathbf{t}_{\mathrm{obs}}$, does not explicitly depend on ${\bf v}$.  
Thus, $\nabla_{\bf v} \log p({\bf v}|{\bf T}_{\rm obs})$ cannot be calculated directly.  
On the other hand, when $\mathbf{t}_{\mathrm{calc}}$ satisfies the eikonal equation for the given velocity structure, as realized by the PINN-based approach in Equation \ref{eqn:argmin_theta_T}, $\mathbf{t}_{\mathrm{calc}}$ implicitly depends on ${\bf v}$.  
In such cases, one can calculate the gradient by adding a Lagrange multiplier term to the negative log-PDF.  
This is a discrete version of the adjoint method \cite[]{Lewis1985}.  
See Appendix \ref{sct:appendix_adjoint} for details of the adjoint-based calculation of the gradient.

\begin{figure*}
\begin{center}
\begin{small}
\includegraphics[clip, width=17cm, bb = 0 0 614 271]{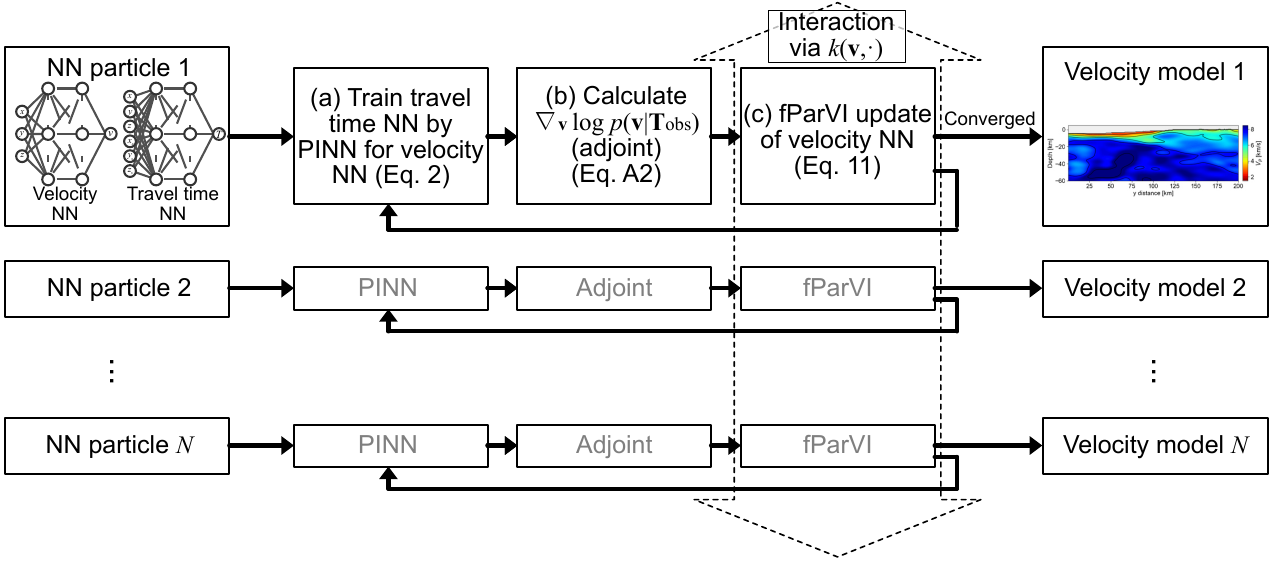}
\end{small}
\end{center}
\caption{Schematic flow of the proposed PINN-based Bayesian travel-time tomography using fParVI.}
\label{fig:fParVI}
\end{figure*}

Here, we summarize the flow of PINN-based Bayesian seismic tomography based on fParVI (Figure \ref{fig:fParVI}).  
Initially, we generate a set of pairs of parameters of the velocity NN $\boldsymbol{\theta}_{v}$ and the travel-time NN $\boldsymbol{\theta}_{T}$, which serve as ``particles'' in ParVI.  
We applied a widely used initialization method to the NN parameters, namely He's initialization \cite[]{He2015}.   
At each iteration, we first train each travel-time NN for the velocity structure given by the partner velocity NN through the PINN framework (Equation \ref{eqn:argmin_theta_T} and Figure \ref{fig:fParVI}(a)).  
Then, we formulate and solve the adjoint equation to obtain the gradient of the negative log-posterior PDF for the velocity structure (Equation \ref{eqn:gradient_adjoint} and Figure \ref{fig:fParVI}(b)).  
Finally, we update the parameters of the velocity NN by fParVI (Equation \ref{eqn:SVGD_update_improved2} and Figure \ref{fig:fParVI}(c)).  
This procedure, which is repeated until the parameters of the velocity NN converge, is independent for each particle except for the computation of the kernel function in fParVI.  
This allows parallel computation to be performed efficiently.  
The posterior PDF of the velocity structure is represented by the ensemble output from the velocity NNs.

Although the procedure presented above, which was first applied to 1D and 2D problems \cite[]{Agata2023}, can be straightforwardly extended to 3D problems in theory, we expect a larger number of evaluation and collocation points $N_v$ and $N_{\rm c}$ to be required for 3D problems, i.e., tens of thousands or more.  
This naturally affects the computational cost, and particular attention must be paid to the choice of the prior distribution to maintain computational scalability.  
In previous studies \cite[]{Agata2023,Agata2025SR}, a Gaussian process with a radial basis function (RBF) kernel was used to define $p({\bf v})$.  
However, the RBF kernel results in a dense covariance matrix, which does not scale well to 3D problems with a larger number of sampling points.  
Instead, we adopted a sparse kernel named the Wendland C4 kernel function \cite[]{Wendland1995ACM}, whose kernel value becomes exactly zero beyond a certain distance.  
This feature gives the covariance matrix a sparse structure, which enables us to introduce a larger number of sampling points.

\subsection{Accounting for uncertainty of passive-source parameters by marginalizing posterior PDF and relocation as a post-process}

\label{sct:marginalization}

Here, we consider the case where the parameters, i.e., location and origin time, of some seismic sources, which correspond to passive sources in practice, are uncertain.
We start by modifying the likelihood function to introduce the shift of the hypocenter location and the origin time of the event, as
\begin{multline}
p(\mathbf{t}_{\mathrm{obs},i} | \mathbf{t}_{\mathrm{calc}}(\mathbf{x}_{\mathrm{s},i}^\mathrm{ref}+\delta \mathbf{x}_{\mathrm{s},i}),\delta t_{\text{ori},i}) \\
= \frac{1}{\sqrt{(2\pi)^{n_i} \det(\mathbf{C}_{\mathrm{data}})}} \exp{\left(-\frac{1}{2} \left(\Delta \mathbf{t}(\mathbf{x}_{\mathrm{s},i}^\mathrm{ref}+\delta \mathbf{x}_{\mathrm{s},i}) -\delta t_{\text{ori},i}*\mathbf{1} \right)^\top \mathbf{C}_{\mathrm{data}}^{-1} \left( \Delta \mathbf{t}(\mathbf{x}_{\mathrm{s},i}^\mathrm{ref}+\delta \mathbf{x}_{\mathrm{s},i}) -\delta t_{\text{ori},i}*\mathbf{1} \right) \right)}, \label{eqn:likelihood_with_origin}
\end{multline} 
where $\delta \mathbf{x}_{\mathrm{s},i}$ and $\delta t_{\text{ori},i}$ are the shifts of the hypocenter location and the origin time of the $i$-th event, respectively.
This equation can be obtained from Equation \ref{eqn:likelihood} by substituting 
\begin{eqnarray}
\mathbf{x}_{\mathrm{s},i} = \mathbf{x}_{\mathrm{s},i}^\mathrm{ref}+\delta \mathbf{x}_{\mathrm{s},i}
\end{eqnarray}
and replacing $\mathbf{t}_\mathrm{calc}(\mathbf{x}_{s,i}) $ with 
  $\mathbf{t}_\mathrm{calc}(\mathbf{x}_{s,i})+\delta t_{\text{ori},i}*\mathbf{1}$. 
The latter leads to replacement of $\Delta \mathbf{t}(\mathbf{x}_{\mathrm{s},i})$ with 
\begin{eqnarray}
\mathbf{t}_{\mathrm{obs},i}-(\mathbf{t}_{\mathrm{calc}}(\mathbf{x}_{\mathrm{s},i})+\delta t_{\text{ori},i}*\mathbf{1})=\Delta \mathbf{t}(\mathbf{x}_{\mathrm{s},i})-\delta t_{\text{ori},i}*\mathbf{1}.
\end{eqnarray}
To simplify the notation, we introduce a new variable $\delta \boldsymbol{\varphi}_i = [\delta \mathbf{x}_{\mathrm{s},i}, \delta t_{\text{ori},i}]^\top$ for the source parameters.
Using this likelihood function, we reformulate the Bayesian estimation and construct the joint posterior PDF of the velocity structure and the source parameters as
\begin{eqnarray}
  p({\bf v}, \delta \boldsymbol{\varphi}_1, \dots, \delta \boldsymbol{\varphi}_N |{\bf T}_{\rm obs}) &\propto&   p({\bf v}) \prod_{i=1}^{N} 
   p(\mathbf{t}_{\mathrm{obs},i} | \mathbf{t}_{\mathrm{calc}}(\mathbf{x}_{\mathrm{s},i}^\mathrm{ref}+\delta \mathbf{x}_{\mathrm{s},i}),\delta t_{\text{ori},i})p(\delta \boldsymbol{\varphi}_{i}).
\end{eqnarray}
Rather than conducting Bayesian estimation directly for the joint posterior PDF, we marginalize over source-parameter uncertainty to obtain the posterior PDF of the velocity structure:
\begin{eqnarray}
p({\bf v}|{\bf T}_{\rm obs}) &=& \int_{\mathcal{\Phi}_{s,1}}\dots\int_{\mathcal{\Phi}_{s,N}} p({\bf v},\delta \boldsymbol{\varphi}_{1},\dots,\delta \boldsymbol{\varphi}_{N}|{\bf T}_{\rm obs}) d\delta \boldsymbol{\varphi}_{1}\dots d\delta \boldsymbol{\varphi}_{N}\\
&\propto&   p({\bf v}) \prod_{i=1}^{N} 
\int_{\mathcal{\Phi}_{s,i}}  p(\mathbf{t}_{\mathrm{obs},i} | \mathbf{t}_{\mathrm{calc}}(\mathbf{x}_{\mathrm{s},i}^\mathrm{ref}+\delta \mathbf{x}_{\mathrm{s},i}),\delta t_{\text{ori},i})p(\delta \boldsymbol{\varphi}_{i}) d\delta \boldsymbol{\varphi}_{i},
\end{eqnarray} 
where $\delta \boldsymbol{\varphi}_{i} \in \mathcal{\Phi}_{s,i}$. 
Here, we assume that $p({\bf v}|\delta \boldsymbol{\varphi}_{i}) = p({\bf v})$, which is a reasonable assumption in most cases.
However, this integration is not tractable analytically in general.
Under the assumption of a Gaussian likelihood function and a Gaussian prior PDF for source parameters, this integration can be approximated by linearizing the forward modeling of travel time with respect to the source parameters around their prior mean based on a first-order Taylor expansion  \cite[]{Tarantola2005Textbook,Duputel2014}.
Specifically, 
\begin{eqnarray}
I_i &=& \int_{\mathcal{\Phi}_{s,i}}  p(\mathbf{t}_{\mathrm{obs},i} | \mathbf{t}_{\mathrm{calc}}(\mathbf{x}_{\mathrm{s},i}^\mathrm{ref}+\delta \mathbf{x}_{\mathrm{s},i}),\delta t_{\text{ori},i})p(\delta \boldsymbol{\varphi}_{i}) d\delta \boldsymbol{\varphi}_{i}\\
&=& \frac{1}{Z} \int_{\mathcal{\Phi}_{s,i}}   \exp{\left(-\frac{1}{2} \left(\Delta \mathbf{t}(\mathbf{x}_{\mathrm{s},i}^\mathrm{ref}+\delta \mathbf{x}_{\mathrm{s},i}) -\delta t_{\text{ori},i}*\mathbf{1} \right)^\top \mathbf{C}_{\mathrm{data}}^{-1} \left( \Delta \mathbf{t}(\mathbf{x}_{\mathrm{s},i}^\mathrm{ref}+\delta \mathbf{x}_{\mathrm{s},i})-\delta t_{\text{ori},i}*\mathbf{1} \right) \right)}
\exp{\left(-\frac{1}{2}  \delta \boldsymbol{\varphi}_{i} ^\top \mathbf{C}_{\mathrm{s},i}^{-1} \delta \boldsymbol{\varphi}_{i} \right) }  d\delta \boldsymbol{\varphi}_{i}\\
&\approx& \frac{1}{Z} \int_{\mathcal{\Phi}_{s,i}}  \exp{\left(-\frac{1}{2} \left( \Delta \mathbf{t}(\mathbf{x}_{\mathrm{s},i}^\mathrm{ref})-\mathbf{G}_i \delta \boldsymbol{\varphi}_{i} \right)^\top \mathbf{C}_{\mathrm{data}}^{-1} \left( \Delta \mathbf{t}(\mathbf{x}_{\mathrm{s},i}^\mathrm{ref})-\mathbf{G}_i \delta \boldsymbol{\varphi}_{i} \right)
 -\frac{1}{2} \delta \boldsymbol{\varphi}_{i}^\top \mathbf{C}_{\mathrm{s},i}^{-1} \delta \boldsymbol{\varphi}_{i} \right) } d\delta \boldsymbol{\varphi}_{i}, \label{eqn:integration_likelihood_approx}
\end{eqnarray}
where 
\begin{eqnarray}
\mathbf{t}_{\mathrm{calc}}(\mathbf{x}_{\mathrm{s},i}^\mathrm{ref}+\delta \mathbf{x}_{\mathrm{s},i}) +\delta t_{\text{ori},i}*\mathbf{1} \approx \mathbf{t}_{\mathrm{calc}}(\mathbf{x}_{\mathrm{s},i}^\mathrm{ref})+\mathbf{G}_i \delta \boldsymbol{\varphi}_{i}. \label{eqn:linearized_prediction}
\end{eqnarray}
$\mathbf{C}_{\mathrm{s},i}$ is the covariance matrix of the prior PDF for the source parameters and $\displaystyle\mathbf{G}_i=\frac{\partial (\mathbf{t}_{\mathrm{calc}}+\delta t_{\text{ori},i}*\mathbf{1})}{\partial \delta\boldsymbol{\varphi}_{i}}|_{\delta \boldsymbol{\varphi}_{i}=\mathbf{0}}$.
Because $\mathbf{C}_{\mathrm{s},i}$ is a small matrix with dimension of 4$\times$4, we can easily incorporate covariance components, which is essential when coordinate-system conversion is considered.
Following the formula for the Gaussian integral, we can calculate the integral analytically as 
\begin{eqnarray}
I_i &\approx& \frac{1}{Z'} \exp{\left(-\frac{1}{2}  \Delta \mathbf{t}(\mathbf{x}_{\mathrm{s},i}^\mathrm{ref})^\top (\mathbf{C}_{\mathrm{data}}+\mathbf{G}_i \mathbf{C}_{\mathrm{s},i} \mathbf{G}_i^\top)^{-1} \Delta \mathbf{t}(\mathbf{x}_{\mathrm{s},i}^\mathrm{ref}) \right)}, \label{eqn:integration_likelihood_approx_2}
\end{eqnarray}
where $Z'$ is a new normalization constant.
In the context of travel-time tomography, this likelihood function is equivalent to Equation \ref{eqn:likelihood} in the case of well-known source locations, taking the prior mean of uncertain source locations as the fixed sources and replacing the data covariance matrix $\mathbf{C}_{\mathrm{data}}$ with $\mathbf{C}_{\mathrm{data}}+\mathbf{G}_i \mathbf{C}_{\mathrm{s},i} \mathbf{G}_i^\top$.
This approximation means that the source parameter uncertainty is incorporated into the likelihood function through the modification of the data covariance matrix.
This modification introduces covariance components into each source-wise data covariance matrix, making it fully dense.
This does not cause a serious computational cost issue if the number of observation points for each source is moderate, i.e., several hundred.
The linear approximation used here should be understood as a local approximation of the calculated travel times with respect to the source parameters for each fixed velocity model. It is expected to be reasonable when the uncertainty in hypocenter location and origin time is sufficiently small that the travel-time changes over this uncertainty range are well represented by the local travel-time sensitivities to these parameters.
This is a simple yet robust approximation that can incorporate the effect of nuisance parameter uncertainty into the likelihood function. 
Although this is not a new approach, recently developed machine-learning libraries, such as PyTorch and JAX, allow us to obtain the gradient of the forward modeling with respect to the nuisance parameters $\mathbf{G}_i$ far more straightforwardly, whereas this previously required analytical expressions or finite-difference approximation. Although the integration can be approximated more accurately by using Monte Carlo integration \cite[]{Gesret2015,Agata2021}, the computational cost becomes prohibitively high for a large number of source locations.
Thus, we obtain the updated posterior PDF of the velocity structure as 
\begin{eqnarray}
p({\bf v}|{\bf T}_{\rm obs}) &\propto&  p({\bf v}) \prod_{i=1}^{N} I_i. 
\end{eqnarray}
The PINN-based Bayesian tomography with fParVI can be straightforwardly applied to this updated posterior PDF to obtain the ensemble of velocity structures.
Active and passive sources,  whose locations and origin times are well known and uncertain, respectively, can be handled simultaneously by simply using the original data covariance matrix for the former and the modified one for the latter. 

We further aim to obtain the posterior PDF of $\delta \boldsymbol{\varphi}_i$ for the $i$-th source and estimate the source parameters of each source as the posterior mean.
Starting from Bayes' theorem, we marginalize over the velocity structure $\mathbf{v}$:
\begin{equation}
p(\delta \boldsymbol{\varphi}_i | T_{\mathrm{obs}})
=
\int p(\delta \boldsymbol{\varphi}_i, \mathbf{v} | {\bf T}_{\rm obs}) d\mathbf{v}
\label{eq:varphi_marginal_1}
\end{equation}
\begin{equation}
=
\int p(\delta \boldsymbol{\varphi}_i | \mathbf{v}, {\bf T}_{\rm obs})\, p(\mathbf{v} | {\bf T}_{\rm obs})\, d\mathbf{v}.
\label{eq:varphi_marginal_2}
\end{equation}
Because samples from $p(\mathbf{v} | {\bf T}_{\rm obs})$ are available through the fParVI-PINN framework described in Section \ref{sct:fParVI}, we can approximate the integration in Equation \ref{eq:varphi_marginal_2} using Monte Carlo integration \cite[]{Agata2021}.
Introducing the same Laplace approximation as in Equation \ref{eqn:integration_likelihood_approx_2} to the integrand, this integration is approximated by a summation of Gaussian PDFs, for which the posterior mean and covariance matrix are calculated in closed form. 
See Appendix \ref{sct:appendix_postprocessing} for details. 

\section{Numerical experiments}
\label{sct:verification}

To validate the ability of the proposed method to perform Bayesian 3D travel-time tomography using active and local passive sources, we applied the proposed method to a synthetic test. 

\subsection{Problem and analysis settings}

\begin{figure*}
\begin{center}
\begin{small}
\includegraphics[clip, width=17cm, bb = 0 0 825 316]{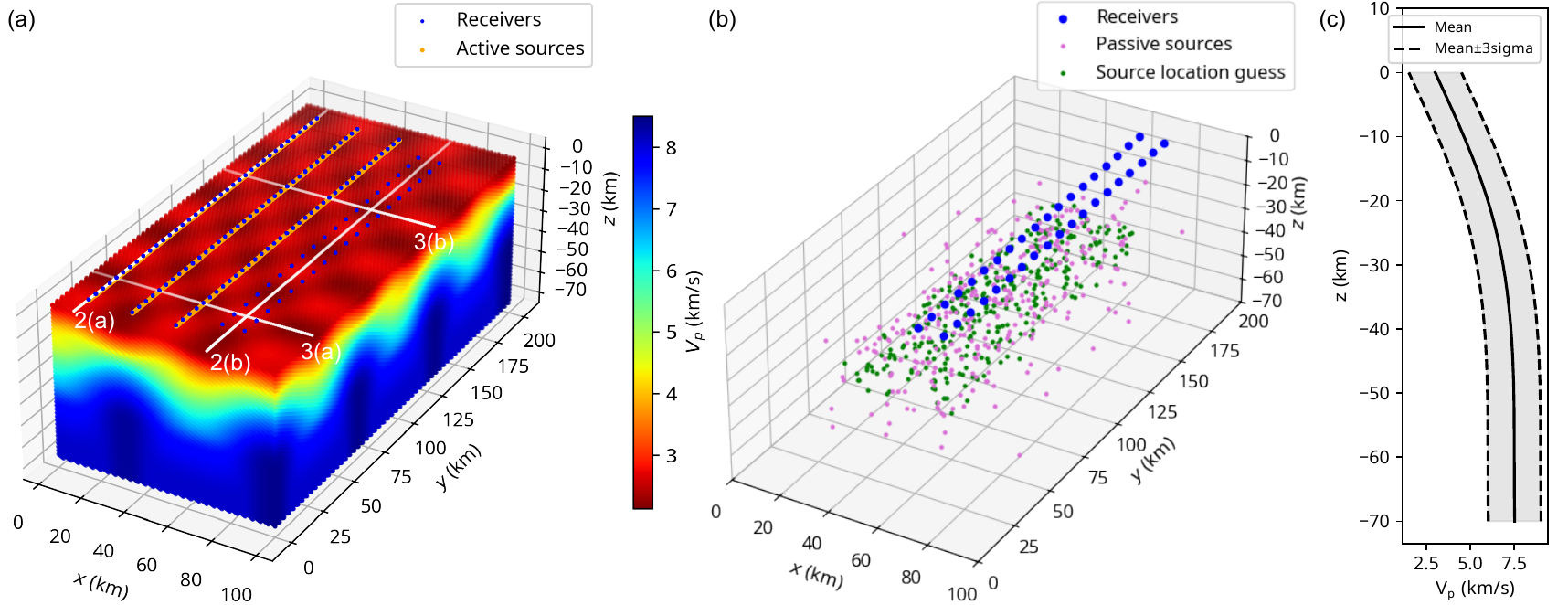}
\end{small}
\end{center}
\caption{Analysis setting for the synthetic numerical experiment. (a) True 3D $P$-wave velocity structure used to generate the synthetic travel-time data and locations of active sources and their receivers. (b) Distribution of passive sources, their receivers, and incorrect source-location guesses adopted in Case 2. (c) The mean and $\pm 3-\sigma$ range of the prior velocity distribution.}
\label{fig:synthetic_velocity_structure}
\end{figure*}

We used a true velocity structure spanning 100\,km, 200\,km, and 70\,km in the x-, y-, and z-directions, respectively, (Figure \ref{fig:synthetic_velocity_structure}(a)).
The background velocity monotonically increases with depth, including sinusoidal structures of shorter and longer wavelength components embedded in the shallower and deeper parts of the velocity structure, respectively.
A total of 543 active sources and 111 receivers were employed at 1-km and 5-km spacing along each of $x=$10, 30, and 50\,km, respectively.
300 passive sources were randomly generated in the domain of $30\,\mathrm{km}\leq x \leq 70\,\mathrm{km}$, $30\,\mathrm{km}\leq y \leq 170\,\mathrm{km}$, and $-41\,\mathrm{km}\leq z \leq -29\,\mathrm{km}$ (Figure \ref{fig:synthetic_velocity_structure}(b)).  
A total of 38 passive-source receivers were employed at 10-km spacing along each of $x=$65 and 75\,km, respectively.
To obtain the synthetic data, we calculated the travel time for each source-receiver pair using the fast marching method \cite[]{Sethian1996,Ganster2023eikonalfm}.
We added Gaussian noise with a variance of $0.1^2$\,s to the synthetic travel-time data. 
We also applied this value to the diagonal components of $\mathbf{C}_\mathrm{data}$ in Equation \ref{eqn:integration_likelihood_approx_2}.
We considered two scenarios for the passive sources:
In Case 1, we used the true source locations as the reference passive source parameters $\mathbf{x}_{{\rm s},i}^\mathrm{ref}$. The purpose here was to validate the ability of the proposed method to recover the true velocity structure in an ideal situation. 
In Case 2, we used $\mathbf{x}_{{\rm s},i}^\mathrm{ref}$ that deviated significantly from the true locations and added deviations to the travel-time data.
These deviated parameters correspond to initial guesses typically obtained from routinely determined solutions or catalogues, simulating a more realistic situation in which the source parameters are uncertain.
In both cases, we specified the standard deviation of the source location prior as 10\,km in the $x$-, $y$-, and $z$-directions. That of the origin time was set to $10/v_\mathrm{prior}(\mathbf{x}_{{\rm s},i}^\mathrm{ref})$\,s, where $v_\mathrm{prior}$ is the mean function of the prior PDF of the velocity structure, as described later. 
This setting corresponds to an uncertainty of the order of 1--3\,s and is relatively conservative compared with the JMA general criterion for well-determined hypocenters, which requires origin-time errors of less than 1.0\,s \cite[]{JMA_catalog_userguide}.
These standard deviations were used to construct $\mathbf{C}_{\mathrm{s},i}$.
The deviations of the source locations and the corresponding travel-time data for the initial guesses in Case 2 were set within the specified uncertainty ranges of the true source parameters.
Note that the travel times for the passive sources were calculated using the true source locations.
To confirm the validity of the marginalization of the posterior PDF over source locations, we also compared a reference case of Case 2 where all the passive sources were fixed to $\mathbf{x}_{{\rm s},i}^\mathrm{ref}$, i.e., the wrong guessed locations.

We adopted a Gaussian process as the prior probability distribution of the velocity structure, defined by the Wendland kernel functions.
The mean velocity was a simple 1D velocity structure (Figure \ref{fig:synthetic_velocity_structure}(c)).  
The standard deviation of the prior distribution was set to 0.5\,km/s.
This choice is conservative, making the 2-$\sigma$ range of the velocity around 1\,km/s.
The scale parameters of the Wendland kernel determine the strength of the correlation embedded in the prior covariance matrix, which controls the degree of smoothness imposed on the velocity structure.
Because the true velocity structure is smoother in the horizontal direction than in the vertical direction, we fixed the ratio of the correlation length in the horizontal direction to that in the vertical direction to 2. 
Following an empirical Bayesian approach, the scale parameter was objectively determined by minimizing the widely applicable Bayesian information criterion (WBIC) \cite[]{Watanabe2013}, which was evaluated across several estimation runs with different scale parameters.
The calculation of WBIC requires the number of independent data points. In this study, we account for the covariance among data components from the same source, which arises from the uncertainty in passive source parameters. Therefore, each data block associated with a passive source is counted as a single independent data point.
For ease of implementation, we conducted these test runs for Case 1 and applied the selected scale parameter to the other cases as well. 
After preliminary runs, we found that fSVGD sometimes underfits the data as the number of particles increases, probably due to the smoothing of the log-posterior gradient. 
In the following analyses presented in this study, we adopted another variant of ParVI, called Gradient Flow with Smoothed Test Functions \cite[GFSF,][]{Liu2019ICML}, instead of SVGD to avoid underfitting \cite[see][for fGFSF: function-space GFSF]{Wang2019}. 
The number of fParVI particles employed for the analyses to obtain the main results was 256. 
Calculations of WBIC were conducted using 32 particles to save computational resources. 

We used fully connected feed-forward neural networks with three hidden layers and 80 hidden units for the velocity NN.  
We used the Deeper-PINN architecture \cite[]{Jiang2024arXiv} with three connection blocks and 80 hidden units for the travel-time NN. 
The activation functions used in the velocity and travel-time NNs are Mish \cite[]{Misra2019arXiv} and Tanh functions, respectively. 
To make it easier for NNs to capture small-scale features, we embedded random Fourier features (RFF) \cite[]{Tancik2020} before passing the input to both types of NNs. 
The frequency of the RFF was determined to match the scale parameter set in the prior PDF of the velocity structure \cite[]{Agata2026CMAMEinprep}.
The SOAP optimizer \cite[]{Vyas2025ICLR}, a state-of-the-art second-order optimization method highly effective for neural network training, was used to determine $\epsilon_l$ with an initial learning rate of $10^{-2}$. 
20,000 collocation points were randomly selected in each iteration and used for PINN training and velocity evaluation. 
All the travel-time data were used in each iteration, i.e., full-batch, and the number of epochs for iteration $l$ was set to 600.
Each training step of $\boldsymbol{\theta}_{T}$ was conducted using the SOAP optimizer, which has been also reported to be effective for improving the convergence of PINN \cite[]{Wang2025NeurIPS}, for 1,000 epochs with $N_c = 20,000$ and an initial learning rate of $10^{-4}$, taking the previous ParVI iteration result as the initial guess.
The coordinates of the velocity evaluation and collocation points were randomly generated in each epoch within the target domain.

The main analysis employing 256 particles for each case was completed in around 23 hours using 64 NVIDIA GH200 Superchips on the Miyabi supercomputer at the Joint Center for Advanced High Performance Computing (JCAHPC). 
This suggests that the proposed PINN-based Bayesian tomography using fParVI can be applied to 3D problems with an acceptable computational cost.

\subsection{Results}

\begin{figure*}
\begin{center}
\begin{small}
\includegraphics[clip, width=17cm]{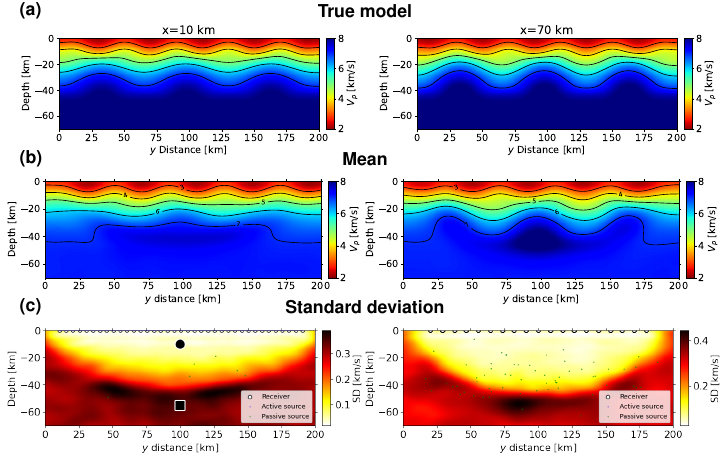}
\end{small}
\end{center}
\caption{Cross-sections of the synthetic numerical experiment Case 1 in the $x$-direction at $x=10$ and 70\,km. From top to bottom, the panels show (a) the true model, (b) the ensemble mean of the proposed method, and (c) the ensemble standard deviation of the proposed method.}
\label{fig:result_NE_x_case1}
\end{figure*}

\begin{figure}
\begin{center}
\begin{small}
\includegraphics[clip, width=8.5cm]{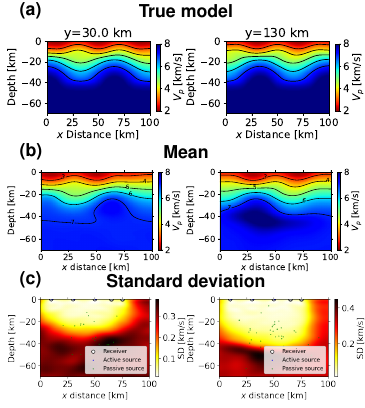}
\end{small}
\end{center}
\caption{Cross-sections of the synthetic numerical experiment Case 1 in the $y$-direction at $y=30$ and 130\,km. The panel arrangement is the same as in Figure \protect\ref{fig:result_NE_x_case1}.}
\label{fig:result_NE_y_case1}
\end{figure}

\begin{figure*}
\begin{center}
\begin{small}
\includegraphics[clip, width=16cm]{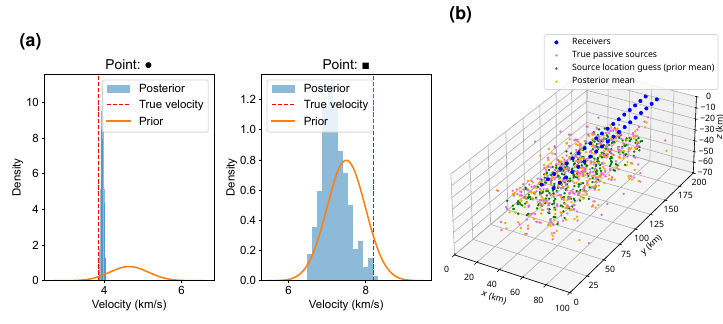}
\end{small}
\end{center}
\caption{(a) Pointwise priors and histograms of the posterior probability density functions of velocity at the three representative points marked in Figure \protect\ref{fig:result_NE_x_case1}(a). The circle and square correspond to the shallow, well-constrained region and the poorly constrained region with large uncertainty, respectively. (b) 3D plot of the true, prior mean, and posterior mean passive-source locations for the synthetic numerical experiment Case 2.}
\label{fig:3D_model_passive_source_posterior}
\end{figure*}

In Case 1, where the prior mean source locations were the same as the true ones, the velocity structure estimated by the proposed method agrees well with the true velocity structure (Figure \ref{fig:result_NE_x_case1}(a)(b) and \ref{fig:result_NE_y_case1}(a)(b)).
The standard deviation across the 256 models can be regarded as an indicator of the estimated uncertainty.
The spatial pattern of the standard deviation is also reasonable, showing large uncertainty in the deep part and in parts distant from the survey lines and smaller uncertainty in the shallower regions that are well covered by the active sources (Figure \ref{fig:result_NE_x_case1}(c) and \ref{fig:result_NE_y_case1}(c)). 
These uncertainty patterns reflect the combined effects of tomographic resolution associated with the uneven ray-path coverage imposed by the source--receiver geometry, data noise including uncertainties in the location and origin time for passive source observations, and the strength of the prior constraint.
These contrasts in the uncertainty patterns between the regions are more apparent in the pointwise visualization of the histogram of the ensembles (Figure \ref{fig:3D_model_passive_source_posterior}(a)).
The shallowest point, indicated by the circle, shows a sharp posterior close to the true value. 
Interestingly, models with velocities larger than 8\,km/s are almost absent even in the region with large uncertainty, indicated by the triangle point, where few rays pass through. 
This is probably because a large velocity here would modify the ray paths to go through this region and strongly affect the travel time, while a small velocity would not.
As a result, the true velocity is located at the edge of the histogram. 
Such a nonlinear effect can be captured in our Bayesian estimation thanks to the ensemble-based approach.
Overall, this result demonstrates the applicability of the proposed method to 3D travel-time tomography in an ideal situation.

\begin{figure*}
\begin{center}
\begin{small}
\includegraphics[clip, width=17cm]{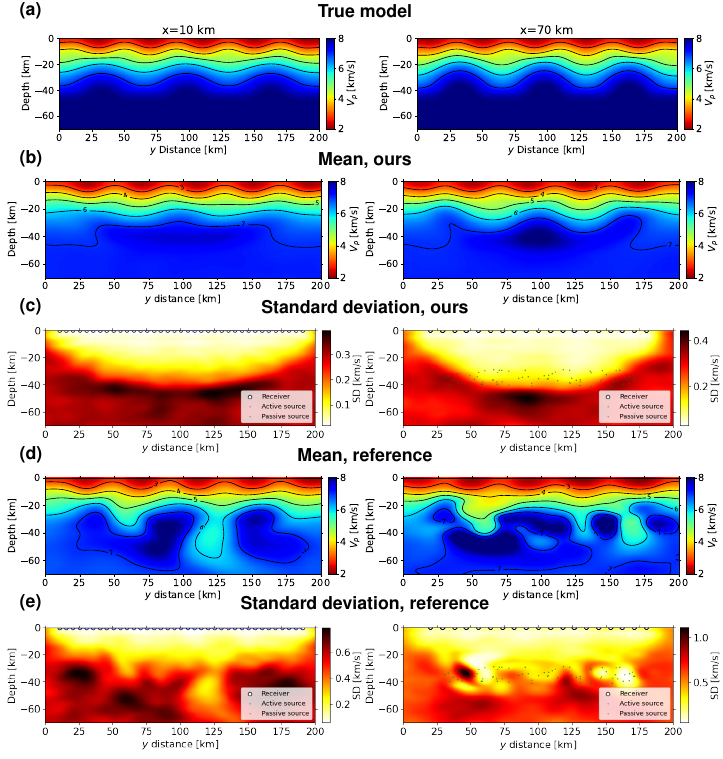}
\end{small}
\end{center}
\caption{Cross-sections of the synthetic numerical experiment Case 2 in the $x$-direction at $x=10$ and 70\,km. From top to bottom, the panels show (a) the true model, (b) the ensemble mean of the proposed method, (c) the ensemble standard deviation of the proposed method, (d) the mean of the reference inversion results with fixed incorrect passive-source locations, and (e) the corresponding standard deviation.}
\label{fig:result_NE_x_case2}
\end{figure*}

\begin{figure}
\begin{center}
\begin{small}
\includegraphics[clip, width=8.5cm]{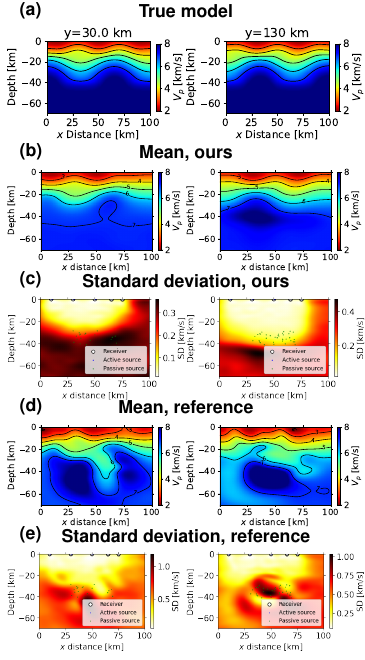}
\end{small}
\end{center}
\caption{Cross-sections of the synthetic numerical experiment Case 2 in the $y$-direction at $y=30$ and 130\,km. The panel arrangement is the same as in Figure \protect\ref{fig:result_NE_x_case2}.}
\label{fig:result_NE_y_case2}
\end{figure}

In Case 2, where the prior mean source locations deviated significantly from the true ones, the estimated mean velocity structures and uncertainty maps agree surprisingly well with those of Case 1 (Figure \ref{fig:result_NE_x_case2}(b)(c) and \ref{fig:result_NE_y_case2}(b)(c)). 
The results suggest that the theoretically justified data covariance matrix introduced in Equation \ref{eqn:integration_likelihood_approx_2} successfully accounts for the uncertainty of the source locations in a statistically consistent manner. 
By contrast, the reference case of Case 2 resulted in a mean velocity structure significantly contaminated by artifacts due to the wrong guesses of the source locations (Figure \ref{fig:result_NE_x_case2}(d) and \ref{fig:result_NE_y_case2}(d)). 
These artifacts and deviations in the velocity structures in the reference case are due to overfitting to the wrongly guessed source locations.
The standard deviation of the results of this case is unnaturally small even in the deep regions of the velocity structures, where the mean models appear to be contaminated (Figure \ref{fig:result_NE_x_case2}(e) and \ref{fig:result_NE_y_case2}(e)). 
Estimates in these regions are likely overconfident due to overfitting.
In addition, the largest value of the standard deviation is 0.8 km/s or larger. This is inconsistent with the prior probability distribution of the velocity structure, whose standard deviation is 0.5 km/s.

In Case 2, we conducted post-processing to obtain the posterior PDF of the source parameters following the procedure described in Appendix \ref{sct:appendix_postprocessing}. 
Here, we only present the posterior mean. 
The posterior mean of the source location is significantly closer to the true location than the prior mean (Figure \ref{fig:3D_model_passive_source_posterior}(b)).
We quantify this improvement in location using the root-mean-square (RMS) error relative to the true source locations as follows:
\begin{equation}
\mathrm{RMS}_\mathrm{source} = 
\sqrt{\frac{1}{N_s}\sum_{i=1}^{N_s}\left|\mathbf{x}_{s,i}-\mathbf{x}_{s,i}^{\mathrm{true}}\right|_2},
\end{equation}
where $N_s$ is the number of sources and $\mathbf{x}_{s,i}^{\mathrm{true}}$ is the true location of the $i$-th source. 
In Case 2, the RMS error decreased significantly from 17.4\,km for the prior mean locations to 6.0\,km for the posterior mean locations.
The remaining error is likely due to the ill-conditioned nature of the problem and the effect of the linear approximation involved in the integration of the posterior PDF.

We further compared the root-mean-square (RMS) of the travel-time misfit for four subsets of datasets, i.e., active-source data, passive-source data, and all data, obtained in Case 2 using our approach and the reference one.
The RMS misfit for travel-time data is calculated as follows:
\begin{equation}
\mathrm{RMS}_\mathrm{traveltime}
= \sqrt{\frac{1}{N}\sum_{i=1}^{N}\left(t_{{\rm obs},i}-t_{{\rm calc},i}\right)^2}.
\label{eq:rms_misfit}
\end{equation}
In Case 2 using the proposed method, misfits of passive-source and all data calculated using both the prior and posterior mean of the source parameters are shown (Table \ref{tab:rms_misfit_NE}). 
Because the RMS misfit differs among our 256 models, we show the results for the models with the smallest and median misfit. 
For all models, the misfit for the active source is much smaller than that for the passive source. In the reference case, the misfit is reduced for the incorrectly located sources, which in turn increases the misfit for the true active sources. Such overfitting is not observed in the proposed method. As a result, the overall misfit with respect to the prior is smaller for the proposed method. 
In addition, the misfits for both the passive source and the overall dataset calculated for the posterior mean, i.e. the relocated hypocenters and the estimated origin time shifts, are reduced.
This result demonstrates the ability of the proposed method to mitigate overfitting of the velocity structure to uncertain source parameters and to relocate them to positions closer to the true ones.

\section{Application to estimation of the P-wave velocity structure off the Kii Peninsula using active and local passive sources}

\label{sec:application}
\subsection{Data and analysis setting}

\begin{figure*}
\begin{center}
\begin{small}
\includegraphics[clip, width=17cm, bb = 0 0 935 388]{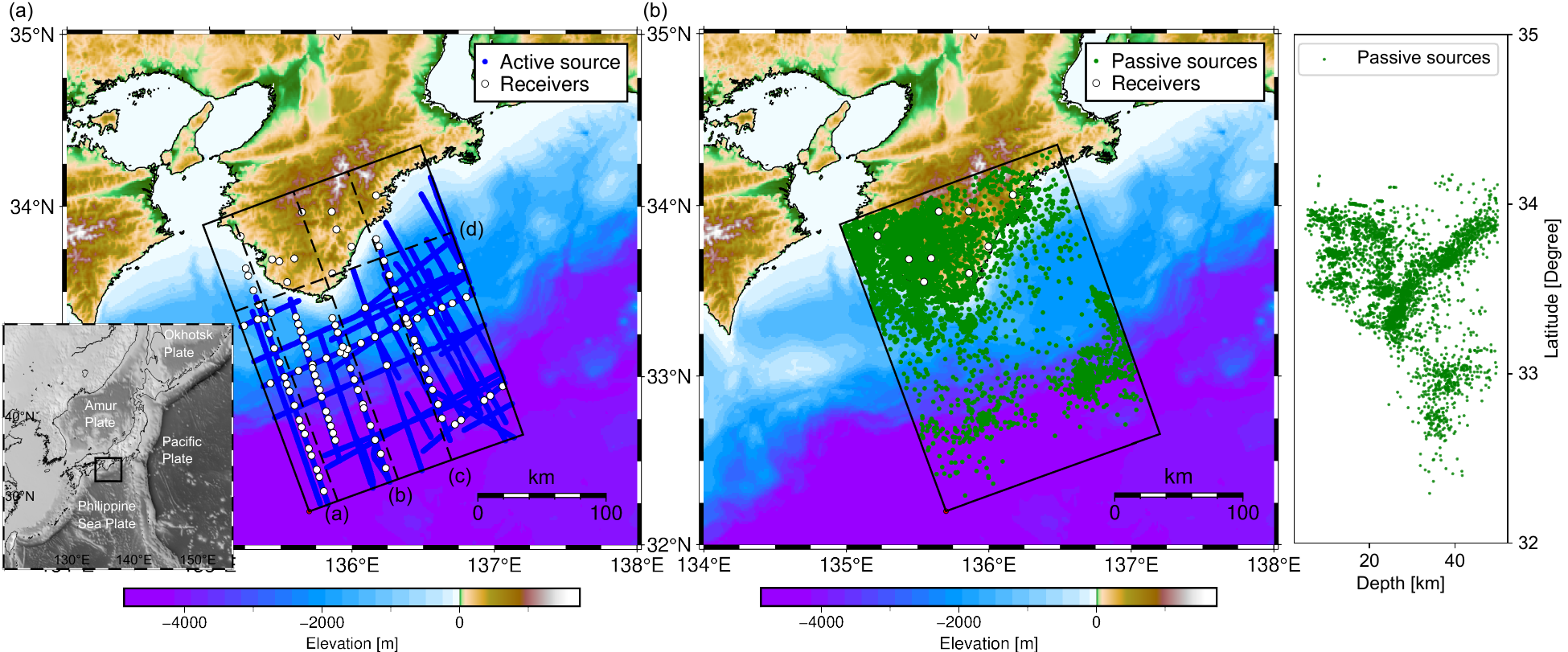}
\end{small}
\end{center}
\caption{Study area and dataset geometry for the application off the Kii Peninsula. (a) The target model domain with the distributions of the active-source shots and receivers. (b) Distribution of passive earthquake sources and their receivers.}
\label{fig:map}
\end{figure*}

The offshore region of the Kii Peninsula in Southwest Japan is located within the Nankai Trough subduction zone, where great earthquakes of M8-9 have repeatedly occurred.
This region includes segments of adjacent M8-class earthquakes that have ruptured in close succession multiple times \cite[]{Ando1975}.
Extensive marine active-source seismic surveys have been conducted in this region \cite[see][]{JAMSTEC2016}, resulting in the accumulation of a large volume of active-source seismic data. 
In addition, because of the high seismicity in this region, first-arrival travel times from natural earthquakes have been widely recorded by onshore permanent seismic networks, making them available as passive-source data.
These first-arrival picks have been compiled and used in previous studies to perform travel-time tomography for the entire Nankai Trough region \cite[]{Arnulf2022NatGeo,Bassett2022JGR-SE}.
In this study, we extracted travel-time data corresponding to the target region off the Kii Peninsula from this dataset and applied the proposed Bayesian travel-time tomography method for P-wave velocity structure estimation.
We targeted a domain located off the Kii Peninsula, Southwest Japan, spanning 150\,km, 200\,km, and 60\,km in the $x$-, $y$-, and $z$-directions, respectively (Figure \ref{fig:map}).
The extracted dataset includes 322,549 and 61,710 travel-time data from 48,134 active sources and 9,107 passive natural earthquake sources, respectively (Figure \ref{fig:map}).
The active-source datasets were acquired by marine seismic surveys using airgun shots and ocean bottom seismometers (OBSs) conducted by JAMSTEC over more than two decades from 1994 \cite[]{JAMSTEC2016}. 
Onshore broadband seismometers, namely F-net and Hi-net operated by the Japanese National Research Institute for Earth Science and Disaster Prevention (NIED), also recorded the onshore arrival of seismic energy originating from the offshore seismic surveys described above (Figure \ref{fig:map}(a)). 
Passive-source datasets were compiled from earthquakes in the catalog of the Japan Meteorological Agency (JMA) that were recorded by Hi-net between 2000 and 2018 (Figure \ref{fig:map}(b)). 
While the original dataset fixed the source locations to the JMA catalogue hypocenters, we considered the uncertainty in the source parameters. 
We assumed that the prior PDF of the source location follows a Gaussian distribution with the mean given by the JMA hypocenters and a standard deviation of 10\,km in the $x$-, $y$-, and $z$-directions.  
This value of the standard deviation was chosen by referring to the relocation result of JMA catalogue hypocenters in the same region using data from a permanent OBS network, i.e., DONET: Dense Oceanfloor Network system for Earthquakes and Tsunamis \cite[]{Kaneda2015,Aoi2020}.
Their result suggests a maximum shift of 10–15\,km shift from the original locations \cite[]{Yamamoto2022PEPS}.
The standard deviation of the origin time was set in the same manner as in the numerical experiments.

To simulate the wavefront propagation from airgun shots to OBSs, the calculation of travel time along the ray paths passing through the interface between the seawater and the solid Earth is required. This interface exhibits a significant velocity discontinuity.
PINNs are known to often have difficulty in accurately learning solutions in cases with sharp parameter discontinuities, as in our case. 
To address this issue and obtain accurate travel-time solutions, we combined PINNs to calculate the travel time for ray paths within the solid Earth and an analytical calculation for ray paths within the seawater by assuming a uniform acoustic wave speed in the seawater. 
See details in Appendix \ref{sct:appendix_traveltime_calculation}.

Similar to the numerical experiments, we adopted a Gaussian process as the prior probability distribution of the velocity structure, defined by the Wendland kernel functions.
The mean velocity was taken from a previously proposed 3D velocity model of the Nankai Trough region, developed by compiling and interpolating the results of 2D tomographic models \cite[]{Nakanishi2018}.
The standard deviation of the prior distribution was set to 0.7\,km/s.
This is a safe choice, making the 2-$\sigma$ range of the velocity 1.4\,km/s.
Determination of the spatial scale parameters of the Wendland kernel function followed the numerical experiments, using WBIC. 
The number of fParVI particles employed here was 255 for the main analyses and 30 for calculations of WBIC.

The NN architecture and training setting here also mostly followed the numerical experiments. 
The number of collocation points for PINN training and evaluation points for the velocity structure was set to 50,000, larger than that in the numerical experiments, to capture the smaller-scale structure of the velocity model.
The batch size of travel-time data was set to approximately one-fifth of the total number, and the number of epochs was set to 120.

\subsection{Results}

\begin{figure*}
\begin{center}
\begin{small}
\includegraphics[clip, width=17cm, bb = 0 0 347 409]{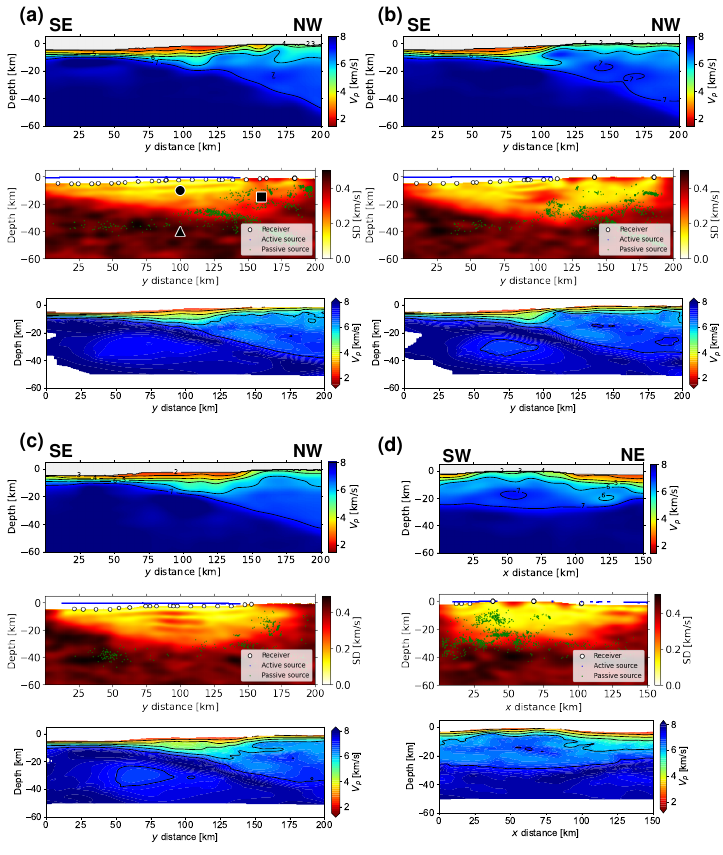}
\end{small}
\end{center}
\caption{Cross-sections of the estimated $P$-wave velocity structure off the Kii Peninsula. Panels (a)--(d) show the sections along $x=20$, 62.5, and 100\,km and along $y=140$\,km, respectively. For each section, the upper, middle, and lower rows show the ensemble mean of our result, the corresponding standard deviation, and the model of \protect\cite{Arnulf2022NatGeo}, respectively. SE, NW, SW, NE indicate the directions of the sections. The circle, square, and triangle in the standard deviation panel in (a) indicate the locations for the velocity histograms shown in Figure \protect\ref{fig:histogram}.}
\label{fig:kii_cross_section}
\end{figure*}

\begin{figure*}
\begin{center}
\begin{small}
\includegraphics[clip, width=15cm, bb = 0 0 845 341]{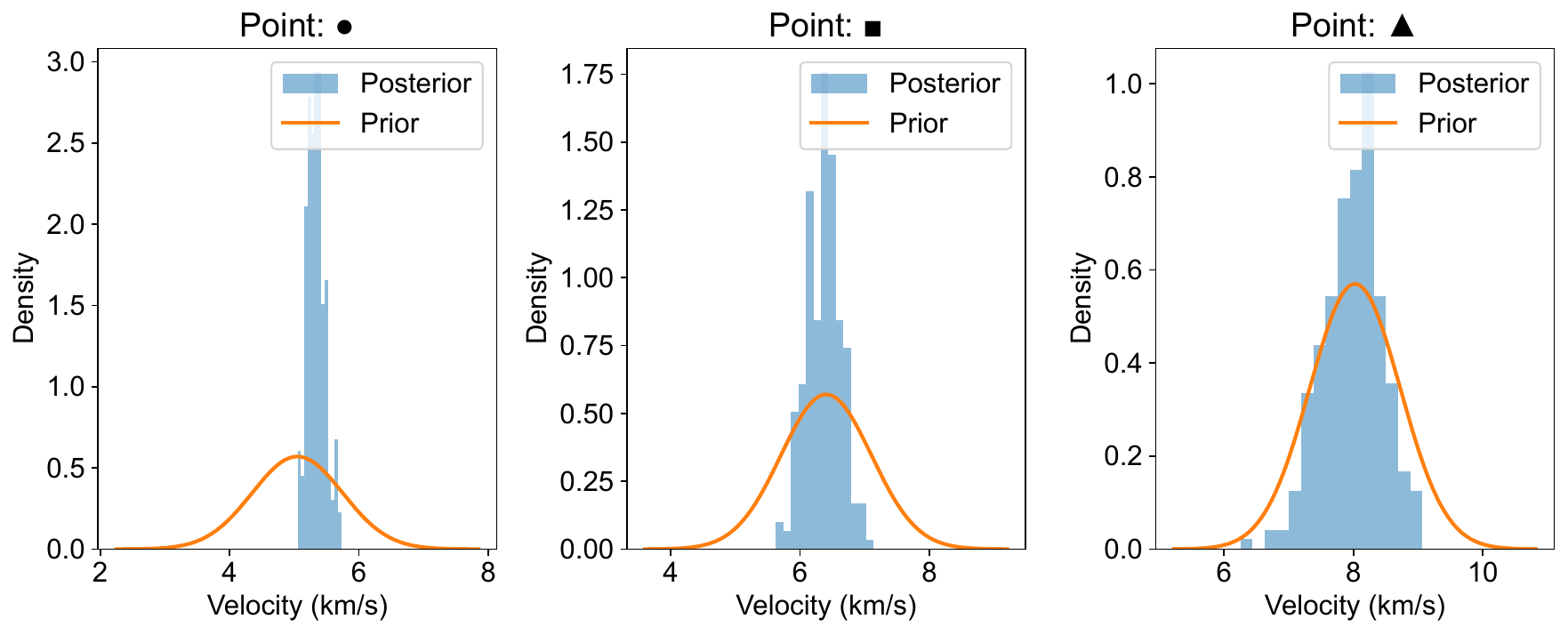}
\end{small}
\end{center}
\caption{Pointwise prior and histogram of the posterior probability density functions of velocity at the three representative points marked in Figure \protect\ref{fig:kii_cross_section}(a). The circle, square, and triangle correspond to the shallow, well-constrained region, the deeper region mainly constrained by passive-source data, and the poorly constrained region with the largest uncertainty, respectively.}
\label{fig:histogram}
\end{figure*}

The cross-sections of the estimated results show that our ensemble-mean model computed from 255 realizations generally agrees well with the results of \cite{Arnulf2022NatGeo} (Figure \ref{fig:kii_cross_section}). 
The 7\,km/s contour dipping toward the northwest shown in Figure \ref{fig:kii_cross_section}(a), (b), and (c) approximately illustrates the overall geometry of the upper surface of the subducting oceanic plate.
Uncertainty is generally small in the shallow portion with dense receiver coverage because of the strong constraints from the active-source data, e.g., the region around the circle in Figure \ref{fig:kii_cross_section}(a). 
Even in the shallow portion, relatively large uncertainty is locally estimated, and its distribution corresponds well to the location of the receiver points.
As depth increases, the constraints from passive-source data of natural earthquakes become dominant, and uncertainty increases, e.g., around the square.
The uncertainty is largest in regions where neither active-source nor passive-source data provide strong constraints, e.g., around the triangle.
Assuming an oceanic crustal thickness of approximately 6 km, the region well constrained by our model can be interpreted as extending down to the oceanic crust, whereas the underlying upper mantle is generally less resolved. 
These contrasts in uncertainty patterns are more apparent in the pointwise visualization of the histogram of the ensembles (Figure \ref{fig:histogram}).
Models with velocities larger than 9\,km/s are almost absent even in the region with the largest uncertainty, i.e., the triangle, where few rays pass through. 
Note that, although a P-wave velocity greater than 8.5\,km/s is generally quite uncommon in subduction zones, the present dataset does not strongly rule out such high velocities.
In the same way as in the numerical experiment, the probability of large velocity in the deep portion tends to become lower than that of small velocity, reflecting the nonlinearity of the inverse problem.

The posterior standard deviation in sections (a)–(c) in Figure \ref{fig:kii_cross_section} exhibits horizontally elongated, stripe-like patterns in the shallow to intermediate depth range. In contrast, such patterns are not clearly observed in section (d), which is not directly overlain by the active-source refraction profile. This contrast suggests that the stripe-like features are related to the active-source acquisition geometry. The patterns also partly coincide with structures in the second vertical derivative of the posterior-mean velocity (Figure S1), suggesting that they may reflect changes in ray sensitivity caused by local ray focusing around the shallow portion. We therefore do not interpret these patterns as those caused by direct geological features.

\begin{figure}
\begin{center}
\begin{small}
\includegraphics[clip, width=8.5cm, bb = 0 0 550 182]{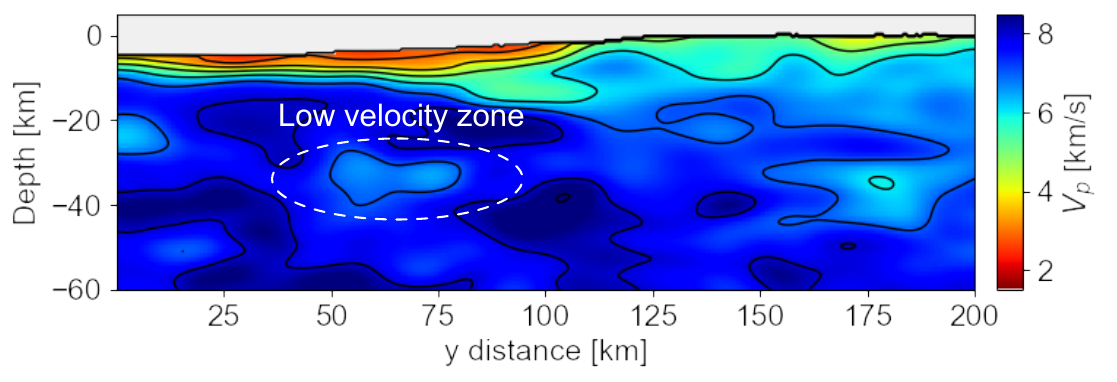}
\end{small}
\end{center}
\caption{Representative posterior sample corresponding to the cross-section of Figure \protect\ref{fig:kii_cross_section}(b), which includes a low-velocity zone in the subducting oceanic plate similar to that seen in \protect\cite{Arnulf2022NatGeo}.}
\label{fig:Low_velocity_sample}
\end{figure}

\begin{figure}
\begin{center}
\begin{small}
\includegraphics[clip, width=8.5cm]{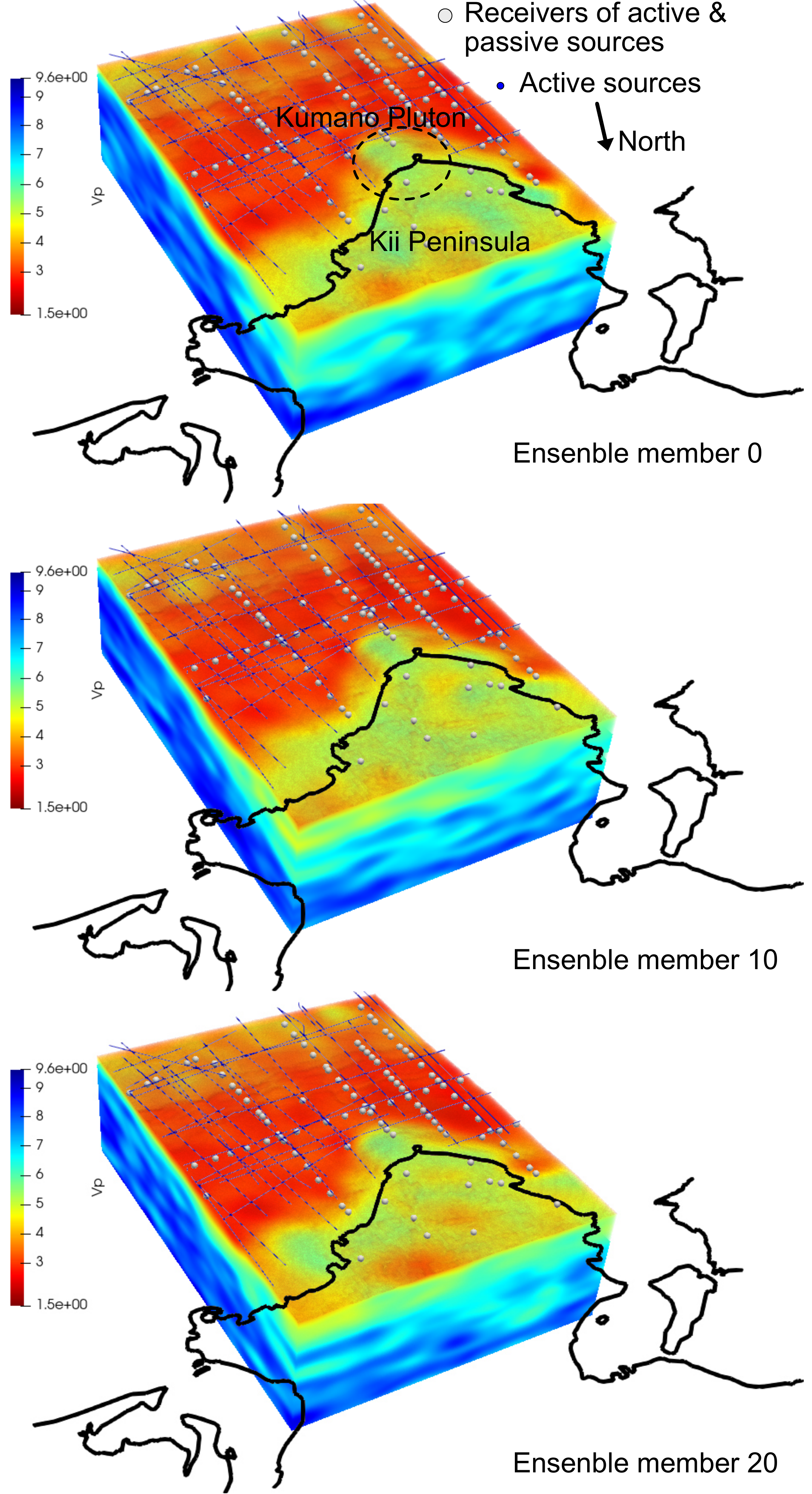}
\end{small}
\end{center}
\caption{Examples of 3D velocity models sampled from the 255-member posterior ensemble for the Kii Peninsula application. The solid black lines denote the coastlines, below which lies the terrestrial region.}
\label{fig:255_ensemble}
\end{figure}

The deviation of the ensemble-mean model from the Arnulf model is observed in the shallowest portion in the terrestrial part (125\,km $\leq y \leq$ 200\,km). This deviation is attributed to the effects of these uncertainty patterns reflecting the spatial sparsity of the receiver points.
The Arnulf model shows a low-velocity zone, partially slower than 7\,km/s, within the subducting oceanic mantle at a depth of 20-40\,km, particularly visible in Figure \ref{fig:kii_cross_section}(b) and (c).  In contrast, a subsequent study \cite[]{Yamamoto2022PEPS}, based on tomographic results obtained by relocating the JMA catalogue hypocenters, does not show such a low-velocity zone.  Although some samples from our velocity ensemble include models that exhibit a similar low-velocity zone (Figure \ref{fig:Low_velocity_sample}), the mean probabilistic model does not contain this feature. Therefore, from a stochastic point of view, the existence of the low-velocity zone is not strongly supported, although considerable uncertainty remains there due to poor raypath coverage.  These interpretations are only possible through the estimation of model uncertainty, highlighting the importance of uncertainty quantification. 
The cross-section on Line (b) (Figure \ref{fig:kii_cross_section}(b)) shows the existence of a high-velocity volume larger than 6\,km/s at the region of $100\,\mathrm{km}\leq y \leq 125\,\mathrm{km}$ and the depth of $5\,\mathrm{km}\leq z \leq 10\,\mathrm{km}$, a structure known as the Kumano Pluton \cite[]{Arnulf2022NatGeo} or the Shionomisaki igneous complex \cite[]{Honda2005EPS,Kodaira2006,Qin2021GRL}, in both our mean model and the Arnulf model.
Visualization of the 3D volume of ensemble members (Figure \ref{fig:255_ensemble} and Supplementary Video) demonstrates that the Kumano Pluton is consistently resolved across the members thanks to intensive seismic exploration over recent decades, while the deeper portion and inland region are less constrained and show large variations among the velocity models. 

\begin{figure*}
\begin{center}
\begin{small}
\includegraphics[clip, width=16cm]{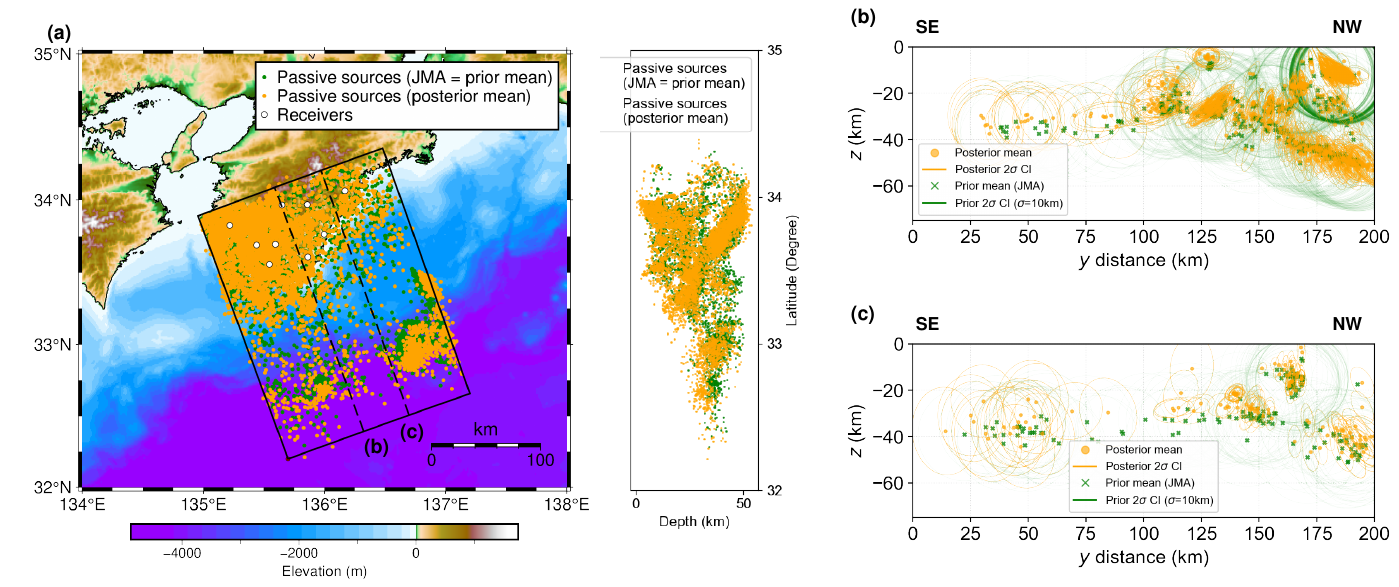}
\end{small}
\end{center}
\caption{Relocation of JMA-catalog hypocenters for passive sources. (a) Map view comparing prior and posterior mean source locations. (b)-(c) Cross-sectional views comparing the posterior mean and standard deviation of source locations. SE and NW indicate the directions of the sections.}
\label{fig:relocation}
\end{figure*}

Relocation of the JMA catalog hypocenters resulted in moderate changes in source locations (Figure \ref{fig:relocation}). The changes were somewhat larger in the depth direction, reaching up to about 10–15\,km, and most events became shallower. This tendency is consistent, although the datasets are different, with the relocation results of \cite{Yamamoto2022PEPS}, which were obtained by double-difference travel-time tomography using permanent ocean bottom seismic stations in this region, including DONET.
The 2-$\sigma$ confidence intervals of the posterior hypocenters suggest that the uncertainty becomes smaller for sources located closer to the onshore region, whereas it remains relatively large for sources located farther offshore (Figure \ref{fig:relocation}(b)(c)).
This contrast likely reflects differences in source--station geometry, particularly the distance from each source to the onshore stations. In addition, spatial variations in the uncertainty of the estimated velocity structure may contribute to the hypocenter uncertainty, because velocity uncertainties are propagated into the travel-time predictions used for source relocation.
In addition, the uniform prior standard deviation of 10\,km assigned to all source locations may itself have been an overestimation for events occurring closer to the land area. 
Note that, because the prior standard deviation of the source locations was initially set to 10\,km, the relocation is in principle constrained to remain within this order of magnitude even if the data could allow a large shift.

To validate the ability of our model to explain the observed travel-time data, we compared the RMS of the travel-time misfit for different data subsets in the same manner as in the numerical experiments (Table \ref{tab:rms_misfit_real}).
For the active-source data, the RMS misfit in our model is smaller than that in the Arnulf model.
That for the passive-source data with prior mean source parameters is significantly larger in our model, leading to a larger RMS misfit for all data as well. 
This contrast reflects the difference in the weights that the two models assigned to each dataset: our model automatically applied a smaller effective weight to passive-source data by introducing a dense covariance matrix theoretically derived from the source parameter uncertainty. 
The same quantity becomes significantly smaller when calculated for the posterior mean, i.e., relocated source locations, but it is still larger than that in the Arnulf model.
As a result, the total RMS for the posterior mean source parameters in our model is smaller than that for the JMA sources in the Arnulf model. 
The JMA source parameters were obtained by assuming a 1D velocity structure. 
Correcting the parameters through the tomographic analysis decreased the total RMS, resulting in both mitigation of overfitting to uncertain passive-source parameters and improved compatibility between the model and the data.

\section{Discussion}

We present a 3D Bayesian tomography result with UQ for the P-wave velocity structure of a subduction zone, yielding a posterior ensemble of physically interpretable 3D velocity models rather than a single best-fit image. 
While Bayesian 3D tomography has been explored in a limited number of studies, many of them rely on sparse grid-based parameterizations, such as Voronoi cells, of the velocity field.
In contrast, our approach explores the posterior in a continuous 3D representation by parameterizing the velocity structure with a neural network. 
A Monte Carlo (MC) analysis with randomized initial models has often been used to assess the stability of solutions obtained by conventional travel-time tomography \cite[]{Korenaga2000,Kodaira2014}. 
This practice reflects the fact that, in many nonlinear inversions, the recovered model can be sensitive to the initial model and other implementation choices, e.g., regularization and hyperparameters. 
By contrast, Bayesian tomography formulates uncertainty through an explicit prior and a data-error model, yielding a posterior distribution that is primarily controlled by the information content of the observations. 
Hence, Bayesian UQ is less dependent on initialization and more directly reflects the quality and quantity of the data. 
The resulting posterior UQ provides a quantitative, data-consistent characterization of spatial variability and model reliability. 

In this study, we implicitly accounted for the uncertainty of passive-source parameters in the posterior estimation of the velocity structure by marginalizing them out, and then performed source relocation as a post-processing step by estimating the posterior distributions of the source parameters. This strategy enables us to obtain the marginal posterior distributions of both the velocity structure and the source parameters, while avoiding the computationally challenging problem of simultaneously estimating them in a single joint Bayesian inversion. 
This marginalization technique has been used in several studies as a means of eliminating nuisance parameters in Bayesian inference \cite[]{Tarantola2005Textbook,Duputel2014,Gesret2015,Agata2021}. However, attempts to recover the posterior distributions of the parameters that have been marginalized out and to re-estimate them after the main posterior inference are still limited \cite[]{Agata2021}. 
In particular, to the best of our knowledge, this study provides the first formulation of such post-processing estimation for the more widely used analytical marginalization method based on linearization of the forward problem and the Laplace approximation \cite[]{Tarantola2005Textbook,Duputel2014}.
Because similar requirements and difficulties for joint estimation arise in various geophysical inverse problems, the proposed approach is expected to have broad applicability.

We point out two important features of the neural representation of the velocity structure in Bayesian travel-time tomography.
One is data efficiency, which enables us to handle ensemble modeling of the 3D velocity structure easily. 
The data size of the neural network employed in this study to model the velocity structure off the Kii Peninsula was less than 8\,MB.  
Even when considering 255 particles for the ensemble estimation, the total data size was still less than 2\,GB. 
This data amount is much smaller than that required for the grid-based representation: with a grid size of 0.5\,km, the data size of a single P-wave velocity model of the target region would be 461\,MB, which would result in more than 100\,GB of data for the ensemble estimation.
This large reduction in data size motivates us to consider applying the proposed method to ensemble modeling of larger-scale velocity structures, such as those of the entire subduction zone.
The other is the mesh-free nature, which allows an infinite-dimensional representation of the velocity structure.
In the real-world application to the region off the Kii Peninsula in this study, this property was particularly essential for the incorporation of bathymetry, serving as the boundary between the seawater and the solid Earth, in the velocity model. 
The mesh-free nature of the proposed method allows the bathymetry to be set at the exact depth taken from the bathymetric data in the model, free from the tradeoff between bathymetry accuracy and grid size. 

Since PINN-based Bayesian tomography relies on deep learning, the use of GPUs is essential for computational efficiency. Although modern GPUs have evolved significantly, memory limitations remain a persistent challenge. 
However, assuming that a neural network is handled within a single memory space, the GPU memory capacity is not yet a bottleneck for the problem size addressed in this study. 
We were able to run multiple processes simultaneously, three for the real data analyses in Section \ref{sec:application}, on a single GPU to utilize resources effectively. 
This suggests that larger neural networks can be employed to cover broader physical regions, such as the entire Nankai Trough. 
Moreover, because (f)ParVI is a gradient-based optimization method, it allows the use of mini-batch learning, randomly selecting a subset of the training data at each iteration.
Mini-batch learning would further extend the method's applicability to problems with significantly larger domains and travel-time datasets.

The posterior uncertainty in poorly resolved regions generally increases toward the prior uncertainty. 
However, in our real-world application presented in Section~\ref{sec:application}, the prior standard deviation in these regions, e.g., near the triangle in Figure \ref{fig:kii_cross_section}(a), was set to \(0.7\,\mathrm{km/s}\), whereas the posterior standard deviation estimated from the particles is approximately \(0.5\,\mathrm{km/s}\). 
As discussed previously, this discrepancy can be partially explained by the asymmetric, model-dependent nature of ray coverage: even if a region is not sampled by rays in a reference or posterior-mean model, models with locally higher velocities may bend or extend ray paths into that region and can therefore be disfavored by the data.
At the same time, part of the reduction may also reflect finite-particle effects and the contractive nature of the gradient-flow dynamics in the ParVI framework. 
Increasing the number of particles may be required for further improvement. 
Future extensions of ParVI methods, such as dynamic-weight ParVI~\cite[]{Zhang2022IJCAI} or hybrid-kernel ParVI~\cite[]{Macdonald2025TMLR}, may also further improve the representation of uncertainty and tail probability in poorly resolved regions.

\section{Conclusion}
\label{sct:conclusion}

In this study, we successfully extended a PINN-based Bayesian travel-time tomography framework to estimate 3D seismic $P$-wave velocity structures and their associated uncertainties. 
This provides, to our knowledge, one of the first practical demonstrations of Bayesian UQ for margin-scale 3D travel-time tomography using both active- and passive-source data.
To overcome the computational challenges of integrating massive datasets of local passive earthquakes alongside active-source data, we introduced a practical and highly scalable marginalization approach. 
We effectively accounted for the uncertainties of passive-source parameters and conducted source relocation without inflating the dimensionality of the primary Bayesian inverse problem by using a marginalization technique. 
Numerical experiments confirmed that our proposed method successfully recovered the true 3D velocity structure and prevented overfitting to biased source priors that severely contaminated conventional approaches. 
Application to the Nankai Trough subduction zone off the Kii Peninsula demonstrated the method's capability to handle real-world complexities, including sharp seawater-crust boundaries and highly heterogeneous data coverage. 
The resulting posterior ensemble of 3D models corroborated the presence of the high-velocity Kumano Pluton and provided quantitative, data-driven uncertainty estimates. 
The relocated hypocenters, i.e., the posterior mean source locations, shifted mainly in the vertical direction by 10-15 km, which is consistent with the deterministic relocation results of a previous study.
Furthermore, the neural network-based mesh-free parameterization achieved a substantial reduction in data storage. 
Combined with the parallel scalability of the fParVI algorithm, this substantial data efficiency demonstrates that the proposed methodology is practical and scalable for broader regional-scale 3D Bayesian tomography and simultaneous UQ in complex tectonic settings.

\clearpage

\begin{table}
\caption{RMS misfit for each model and dataset in the synthetic numerical experiment Case 2. The unit is seconds. The cases ``All (prior)'' and ``All (posterior)'' refer to the RMS misfit for all data using the prior and posterior mean of the passive source parameters, respectively.}
\label{tab:rms_misfit_NE}
\begin{tabular}{lccccc}
\hline
Model  & Active & Passive (prior) & All (prior) & Passive (posterior)& All (posterior)\\
\hline
Ours (smallest)         & 0.108 & 2.016 & 0.163 & 0.803 & 0.118 \\
Ours (median)           & 0.110 & 2.014 & 0.164 & 0.840 & 0.121 \\
Reference (smallest)    & 0.166 & 1.898 & 0.202 & - & - \\
Reference (median)      & 0.178 & 1.921 & 0.212 & - & - \\
\hline
\end{tabular}
\end{table}

\begin{table} 
\caption{RMS misfit for each model and dataset in the real-world application to the region off the Kii Peninsula. The unit is seconds. The cases ``All (prior)'' and ``All (posterior)'' refer to the RMS misfit for all data using the prior and posterior mean of the passive source parameters, respectively.}
\label{tab:rms_misfit_real}
\begin{tabular}{lccccc}
\hline
Model  & Active & \shortstack{Passive\\(prior=JMA)} & \shortstack{All\\(prior=JMA)} & \shortstack{Passive\\(posterior)} & \shortstack{All\\(posterior)}\\
\hline
Ours (smallest) & 0.113 & 0.675 & 0.158 & 0.117 & 0.113 \\
Ours (median)   & 0.119 & 0.714 & 0.167 & 0.111 & 0.119\\   
Arnulf et al.   & 0.142 & 0.152 & 0.144 & - & - \\
\hline
\end{tabular}
\end{table}

\clearpage

\clearpage

\appendix

\section{Appendix}

\subsection{Derivation of the gradient of the loss function using the adjoint method}
\label{sct:appendix_adjoint}

As described in Section~\ref{sct:methods}, computing the gradient of the log-posterior PDF with respect to the velocity structure, $\nabla_{\bf v} \log p({\bf v}|{\bf T}_{\rm obs})$, is essential for driving the fParVI-based Bayesian tomography.
This gradient cannot be evaluated directly because the likelihood function, which is based on the misfit between the forward-modeled travel time $\mathbf{t}_{\mathrm{calc}}$ and the observed data ${\bf T}_{\rm obs}$, does not have an explicit dependence on ${\bf v}$.
However, when $\mathbf{t}_{\mathrm{calc}}$ satisfies the eikonal equation for the given velocity structure, as enforced by the PINN-based forward modeling, there exists an implicit dependence of $\mathbf{t}_{\mathrm{calc}}$ on ${\bf v}$ through the governing equation residual ${\bf r}({\bf T})$.
We exploit this implicit dependence by adopting a discrete version of the adjoint method \cite[]{Lewis1985}.
The negative log-posterior PDF is augmented with a Lagrange multiplier term that enforces the eikonal equation residual ${\bf r}({\bf T})$ to vanish:
\begin{eqnarray}
\mathcal{J} = -\log p({\bf v}|{\bf T}_{\rm obs}) + \boldsymbol{\lambda}^{\top} {\bf r}({\bf T})
\label{eqn:adjoint_equation_appendix}
\end{eqnarray}
where $\boldsymbol{\lambda}$ is the Lagrange multiplier and ${\bf r}$ is the residual vector of the governing equation of the forward modeling, i.e., the eikonal equation.
We emphasize here that ${\bf r}$ depends on the travel time solution ${\bf T}$.
Leveraging the fact that ${\bf r}={\bf 0}$ when the forward modeling is solved properly, the gradient of $\mathcal{J}$, which is equivalent to $-\nabla_{\bf v} \log p({\bf v}|{\bf T}_{\rm obs})$, is calculated as
\begin{eqnarray}
\nabla_{\bf v}\mathcal{J}=\boldsymbol{\lambda}^{*\,\top}\frac{\partial {\bf r}}{\partial {\bf v}},
\label{eqn:gradient_adjoint}
\end{eqnarray}
where $\boldsymbol{\lambda}^{*}$ is the solution of the following adjoint equation:
\begin{eqnarray}
-\frac{\partial \log p({\bf v}|{\bf T}_{\rm obs})}{\partial {\bf T}}+\boldsymbol{\lambda}^{\top}\frac{\partial {\bf r}}{\partial {\bf T}}={\bf 0}.
\label{eqn:adjoint_equation}
\end{eqnarray}
This adjoint equation is solved using the Conjugate Gradient Normal Residual method (CGNR) \cite[]{Saad2003}.
$\displaystyle \frac{\partial {\bf r}}{\partial {\bf v}}$ is the linearized operator of the forward modeling. 
The forward modeling is conducted by introducing a PINN-based eikonal solver for the given velocity, which has been well established by studies such as \cite{Smith2021,Waheed2021PINNeik,Grubas2023}.
Applying two kinds of NNs representing the velocity structure and the travel time solution to this Bayesian estimation framework based on fParVI with the adjoint method described above enables us to complete the Bayesian seismic tomography in a fully meshless framework.
When a grid-based numerical eikonal solver, e.g., the fast marching method \cite[]{Sethian1996}, is used instead of PINN, this approach is identical to a discrete version of the gradient calculation in adjoint-state travel time tomography \cite[]{Tong2021JGR-SE}.

\subsection{Details of post-processing to obtain the posterior PDF of the source locations}
\label{sct:appendix_postprocessing}

We aim to obtain the posterior PDF of the source-parameter perturbation
$\delta \boldsymbol{\varphi}_i = (\delta \mathbf{x}_{\mathrm{s},i}, \delta t_{\mathrm{ori},i})^\top$
for the $i$-th source and estimate the source parameters as the posterior mean.
We assume a Gaussian prior for the perturbation,
\begin{equation}
p(\delta\boldsymbol{\varphi}_i)
=
\mathcal{N}
\left(
\delta\boldsymbol{\varphi}_i;\mathbf{0},\mathbf{C}_{\mathrm{s},i}
\right),
\end{equation}
so that the reference source parameters correspond to the prior mean.
Starting from Bayes' theorem, we marginalize over the velocity structure $v$:
\begin{equation}
p(\delta \boldsymbol{\varphi}_i | T_{\mathrm{obs}})
=
\int p(\delta \boldsymbol{\varphi}_i, v | T_{\mathrm{obs}})\, dv
\label{eq:varphi_marginal_1_app}
\end{equation}
\begin{equation}
=
\int p(\delta \boldsymbol{\varphi}_i | v, T_{\mathrm{obs}})\, p(v | T_{\mathrm{obs}})\, dv.
\label{eq:varphi_marginal_2_app}
\end{equation}
Because only the observed travel times associated with the $i$-th source are involved in the likelihood for $\delta \boldsymbol{\varphi}_i$, we have
\begin{equation}
p(\delta \boldsymbol{\varphi}_i | v, T_{\mathrm{obs}})
=
p(\delta \boldsymbol{\varphi}_i | v, \mathbf{t}_{\mathrm{obs},i}).
\end{equation}
For fixed $v$, Bayes' theorem gives
\begin{equation}
p(\delta \boldsymbol{\varphi}_i | v, \mathbf{t}_{\mathrm{obs},i})
=
\frac{
p(\mathbf{t}_{\mathrm{obs},i} | v, \delta \boldsymbol{\varphi}_i)\, p(\delta \boldsymbol{\varphi}_i)
}{
\displaystyle
\int
p(\mathbf{t}_{\mathrm{obs},i} | v, \delta \boldsymbol{\varphi}_i')\,
p(\delta \boldsymbol{\varphi}_i')\,
d\delta \boldsymbol{\varphi}_i'
}.
\label{eq:conditional_varphi}
\end{equation}

Suppose that we have obtained $M$ particles $\{v^{(m)}\}_{m=1}^M$ from $p(v | T_{\mathrm{obs}})$ using the ParVI algorithm.
Treating these particles as an equally weighted empirical approximation of $p(v | T_{\mathrm{obs}})$, the marginalization over $v$ is approximated as
\begin{equation}
p(\delta \boldsymbol{\varphi}_i | T_{\mathrm{obs}})
\approx
\frac{1}{M}
\sum_{m=1}^M
p(\delta \boldsymbol{\varphi}_i | v^{(m)}, \mathbf{t}_{\mathrm{obs},i}).
\label{eq:mc_varphi}
\end{equation}
For each velocity particle $v^{(m)}$, we linearize the predicted arrival times around $\delta\boldsymbol{\varphi}_i=\mathbf{0}$.
Under the linearized prediction provided by Equation \ref{eqn:linearized_prediction} and assuming Gaussian observation noise with covariance $\mathbf{C}_{\mathrm{data}}$, the conditional posterior for fixed $v^{(m)}$ is approximated by a Gaussian:
\begin{equation}
p(\delta \boldsymbol{\varphi}_i | v^{(m)}, \mathbf{t}_{\mathrm{obs},i})
\approx
\mathcal{N}
\big(
\delta \boldsymbol{\varphi}_i;\,
\boldsymbol{\mu}_i^{(m)},
\mathbf{\Sigma}_i^{(m)}
\big),
\label{eq:laplace_varphi}
\end{equation}
where 
\begin{equation}
\mathbf{\Sigma}_i^{(m)}
=
\left[
\mathbf{C}_{\mathrm{s},i}^{-1}
+
\left(\mathbf{G}_i^{(m)}\right)^\top
\mathbf{C}_{\mathrm{data}}^{-1}
\mathbf{G}_i^{(m)}
\right]^{-1},
\label{eq:posterior_cov}
\end{equation}
\begin{equation}
\boldsymbol{\mu}_i^{(m)}
=
\mathbf{\Sigma}_i^{(m)}
\left(\mathbf{G}_i^{(m)}\right)^\top
\mathbf{C}_{\mathrm{data}}^{-1}
\Delta \mathbf{t}_i^{(m)}(\mathbf{x}_{\mathrm{s},i}).
\label{eq:posterior_mean}
\end{equation}
The normalization constant in the conditional posterior is included in the Gaussian normalization of Eq.~(\ref{eq:laplace_varphi}).
Substituting Eq.~(\ref{eq:laplace_varphi}) into Eq.~(\ref{eq:mc_varphi}), we obtain
\begin{equation}
p(\delta \boldsymbol{\varphi}_i | T_{\mathrm{obs}})
\approx
\frac{1}{M}
\sum_{m=1}^M
\mathcal{N}
\big(
\delta \boldsymbol{\varphi}_i;\,
\boldsymbol{\mu}_i^{(m)},
\mathbf{\Sigma}_i^{(m)}
\big).
\label{eq:mixture_varphi}
\end{equation}
Thus, the marginal posterior PDF of the source-parameter perturbation is represented as an equally weighted Gaussian mixture.
Although this mixture is not necessarily Gaussian, its first two moments can be calculated analytically.
The posterior mean of $\delta \boldsymbol{\varphi}_i$ is then given by
\begin{equation}
\mathbb{E}[\delta \boldsymbol{\varphi}_i | T_{\mathrm{obs}}]
=
\int
\delta \boldsymbol{\varphi}_i\,
p(\delta \boldsymbol{\varphi}_i | T_{\mathrm{obs}})
\, d\delta \boldsymbol{\varphi}_i.
\end{equation}
Using Eq.~(\ref{eq:mixture_varphi}),
\begin{equation}
\mathbb{E}[\delta \boldsymbol{\varphi}_i | T_{\mathrm{obs}}]
\approx
\frac{1}{M}
\sum_{m=1}^M
\boldsymbol{\mu}_i^{(m)}.
\label{eq:posterior_mean_mixture}
\end{equation}
The source parameters after relocation are then obtained as
$\hat{\boldsymbol{\varphi}}_i = [\hat{\mathbf{x}}_{\mathrm{s},i}, \hat{t}_{\text{ori},i}]^\top$,
where
$\hat{\mathbf{x}}_{\mathrm{s},i} = \mathbf{x}_{\mathrm{s},i}^{\mathrm{ref}} + \mathbb{E}[\delta \mathbf{x}_{\mathrm{s},i} | T_{\mathrm{obs}}]$
and
$\hat{t}_{\text{ori},i} = t_{\text{ori},i}^{\mathrm{ref}} + \mathbb{E}[\delta t_{\text{ori},i} | T_{\mathrm{obs}}]$.

The posterior covariance of the source-parameter perturbation can also be obtained straightforwardly from the mixture representation derived above.
Since the posterior PDF of $\delta \boldsymbol{\varphi}_i$ is approximated as an equally weighted mixture of Gaussians, the posterior covariance is given by
\begin{equation}
\mathrm{Cov}[\delta \boldsymbol{\varphi}_i | T_{\mathrm{obs}}]
=
\mathbb{E}
\Big[
(\delta \boldsymbol{\varphi}_i - \bar{\delta \boldsymbol{\varphi}}_i)
(\delta \boldsymbol{\varphi}_i - \bar{\delta \boldsymbol{\varphi}}_i)^\top
| T_{\mathrm{obs}}
\Big],
\end{equation}
where
\begin{equation}
\bar{\delta \boldsymbol{\varphi}}_i
=
\mathbb{E}
[\delta \boldsymbol{\varphi}_i | T_{\mathrm{obs}}]
=
\frac{1}{M}
\sum_{m=1}^M
\boldsymbol{\mu}_i^{(m)}.
\end{equation}
Using standard results for Gaussian mixtures, the posterior covariance is decomposed as
\begin{equation}
\mathrm{Cov}[\delta \boldsymbol{\varphi}_i | T_{\mathrm{obs}}]
\approx
\frac{1}{M}
\sum_{m=1}^M
\mathbf{\Sigma}_i^{(m)}
+
\frac{1}{M}
\sum_{m=1}^M
\left(
\boldsymbol{\mu}_i^{(m)}
-
\bar{\delta \boldsymbol{\varphi}}_i
\right)
\left(
\boldsymbol{\mu}_i^{(m)}
-
\bar{\delta \boldsymbol{\varphi}}_i
\right)^\top.
\label{eq:mixture_covariance}
\end{equation}
The first term represents the average conditional posterior uncertainty of the source parameters for fixed velocity structures, reflecting the effects of observation noise and the source-parameter prior.
The second term represents the variability of the conditional posterior means induced by uncertainty in the velocity structure.

\subsection{Details of travel-time calculation using PINN involving the seawater-solid Earth boundary}
\label{sct:appendix_traveltime_calculation}

As described in Section~\ref{sec:application}, the analysis domain for the real-data application encompasses the seafloor, which constitutes a sharp velocity discontinuity between the seawater and the solid Earth.
Because PINNs are known to have difficulty accurately learning solutions in the presence of sharp parameter discontinuities, we restrict PINN training to the solid Earth domain and treat travel times within the seawater analytically, assuming a spatially uniform acoustic wave speed $v_{\rm sw} = 1.5$\,km/s throughout the seawater layer.

We assume that the source $\mathbf{x}_s$ lies within the solid Earth and the receiver $\mathbf{x}_r$ lies in the seawater. 
Because the P-wave velocity in the solid Earth is usually greater than or equal to the acoustic wave speed in the seawater, a ray that exits the solid Earth into the seawater will not curve back toward the solid Earth.
Consequently, the optimal ray path crosses the seafloor at exactly one point, denoted $\mathbf{x}_{\rm int} \in \Gamma$, where $\Gamma$ is the seafloor surface, and then travels in a straight line to $\mathbf{x}_r$ within the seawater.
The total travel time is therefore given by
\begin{equation}
T(\mathbf{x}_s, \mathbf{x}_r) = \min_{\mathbf{x}_{\rm int} \in \Gamma} \left[ T_1(\mathbf{x}_s, \mathbf{x}_{\rm int}) + T_2(\mathbf{x}_{\rm int}, \mathbf{x}_r) \right],
\label{eqn:traveltime_seawater}
\end{equation}
where $T_1(\mathbf{x}_s, \mathbf{x}_{\rm int})$ is the travel time from $\mathbf{x}_s$ to $\mathbf{x}_{\rm int}$ within the solid Earth, as evaluated by the trained PINN, and $T_2(\mathbf{x}_{\rm int}, \mathbf{x}_r) = |\mathbf{x}_r - \mathbf{x}_{\rm int}|/v_{\rm sw}$ is the travel time along the straight-line path through the seawater.
The training domain of the PINN $\Omega$ in Equation \ref{eqn:eikonal} corresponds to the solid Earth here.
The minimization in Equation \ref{eqn:traveltime_seawater} is performed over the horizontal coordinates of $\mathbf{x}_{\rm int}$ on $\Gamma$ using Newton's method.
The stationarity condition 
\begin{equation}
\nabla_{\mathbf{x}_{\rm int}}(T_1 + T_2) = \mathbf{0}
\label{eqn:stationarity_condition_appendix}
\end{equation}
is equivalent to Snell's law at the seafloor interface.
In practice, this iteration typically converges within five steps.
Rapid convergence is achieved because a good initial estimate of $\mathbf{x}_{\rm int}$ is readily available. We initialize $\mathbf{x}_{\rm int}$ as the vertical projection of the receiver $\mathbf{x}_r$ onto the seafloor $\Gamma$, i.e., the point on $\Gamma$ directly beneath $\mathbf{x}_r$.
Our PINN guarantees source--receiver reciprocity by following the implementation of \cite{Grubas2023}.
The case where the source $\mathbf{x}_s$ lies within the seawater and the receiver $\mathbf{x}_r$ lies in the solid Earth can be treated simply by swapping $\mathbf{x}_s$ and $\mathbf{x}_r$ when inputting the coordinates into the trained PINN.

This modification of the travel-time calculation impacts the formulation of the adjoint method presented in Section \ref{sct:appendix_adjoint}. 
The Lagrangian function is reformulated as 
\begin{eqnarray}
\mathcal{J} = -\log p({\bf v}|{\bf T}_{\rm obs}) + \boldsymbol{\lambda}_1^{\top} {\bf r}_1+
\boldsymbol{\lambda}_2^{\top} {\bf r}_2
\end{eqnarray}
where ${\bf r}_1$ is the residual vector of the eikonal equation and $\boldsymbol{\lambda}_1$ is the Lagrange multiplier for this constraint, i.e., the same as $\boldsymbol{\lambda}$ and ${\bf r}$ in Equation \ref{eqn:adjoint_equation_appendix}.
${\bf r}_2$ is the residual vector for the stationarity condition at the seafloor interface (Equation \ref{eqn:stationarity_condition_appendix}), and $\boldsymbol{\lambda}_2$ is the Lagrange multiplier for this constraint. 
The gradient of the negative log-posterior PDF with respect to the velocity model is given by
\begin{eqnarray}
    -\nabla_{\bf v} \log p({\bf v}|{\bf T}_{\rm obs}) &=& \nabla_{\bf v}\mathcal{J}\\
    &=& \frac{d \mathcal{J}}{d {\bf v}} + \frac{\partial \mathcal{J}}{\partial {\boldsymbol \theta}}\frac{d {\boldsymbol \theta}}{d {\bf v}}  + \frac{\partial \mathcal{J}}{\partial {\bf x}_{\rm int}} \frac{d {\bf x}_{\rm int}}{d {\bf v}},
\end{eqnarray}
where
\begin{eqnarray}
    \frac{\partial \mathcal{J}}{\partial {\boldsymbol \theta}} &=& -\frac{\partial \log p({\bf v}|{\bf T}_{\rm obs})}{\partial {\boldsymbol \theta}} + \boldsymbol{\lambda}_1^{\top} \frac{\partial {\bf r}_1}{\partial {\boldsymbol \theta}} + \boldsymbol{\lambda}_2^{\top} \frac{\partial {\bf r}_2}{\partial {\boldsymbol \theta}} = {\bf 0}\\ \label{eqn:adjoint_equation_12}
    \frac{\partial \mathcal{J}}{\partial {\bf x}_{\rm int}} &=& -\frac{\partial \log p({\bf v}|{\bf T}_{\rm obs})}{\partial {\bf x}_{\rm int}} + \boldsymbol{\lambda}_2^{\top} \frac{\partial {\bf r}_2}{\partial {\bf x}_{\rm int}} = {\bf 0}.
\end{eqnarray}
However, in practice, we usually obtain a good initial guess of ${\bf x}_{\rm int}$ by setting the interface point directly beneath the source location at the depth of the seafloor. In that case, we assume that $\displaystyle \frac{\partial \log p({\bf v}|{\bf T}_{\rm obs})}{\partial {\bf x}_{\rm int}} \approx {\bf 0}$ holds approximately, resulting in $\boldsymbol{\lambda}_2 \approx {\bf 0}$.
Then, the adjoint equation eventually reduces to Equation \ref{eqn:adjoint_equation_appendix}.

\section*{Acknowledgments}

    This study was supported by MEXT Project ``Earthquake Disaster Prevention Research Project for Damage Minimization and Rapid Response and Reconstruction from Great Earthquakes such as the Nankai Trough Earthquake'', JSPS KAKENHI Grant Number 24H01042 25K01084 and ERI JURP 2025-B-01 in Earthquake Research Institute, the University of Tokyo. The calculations were carried out using Earth Simulator at JAMSTEC and Miyabi at JCAHPC. AI-assisted tools were used for translation and grammar checking during the preparation of this manuscript.

\clearpage

\section*{Data Availability}

The travel-time pick data used in this study were extracted from previously compiled active- and passive-source datasets for the Nankai Trough region, as described by \cite{Arnulf2022NatGeo} and \cite{Bassett2022JGR-SE}. The previously published 3D P-wave velocity model of \cite{Arnulf2022NatGeo}, used for comparison in this study, is available from the Marine Geoscience Data System (MGDS) at \url{https://www.marine-geo.org/doi/10.26022/IEDA/329655}. The JMA hypocentre catalogue used to define the prior information for passive-source parameters is available from the Japan Meteorological Agency at \url{https://www.data.jma.go.jp/eqev/data/bulletin/hypo_e.html}, and the Hi-net/F-net station metadata are available from the National Research Institute for Earth Science and Disaster Resilience (NIED) at \url{https://hinetwww11.bosai.go.jp/auth/?LANG=en}.

The posterior velocity-structure ensemble generated in this study, including the neural-network model parameters, relocated passive-source parameters, and figure reproduction data, will be deposited in a public repository before publication. The repository DOI will be provided upon acceptance.

\bibliographystyle{gji}
\bibliography{C:/Users/sgbea/Documents/MyLinks/bibtex_agata}

\label{lastpage}
\end{document}